\documentclass[final,10pt,journal,a4paper,twocolumn]{IEEEtran}
\usepackage[T1]{fontenc}
\usepackage[latin9]{inputenc}
\usepackage{color}
\usepackage{textcomp}
\usepackage{amsmath}
\usepackage{amssymb}
\usepackage{graphicx}

\makeatletter

\providecommand{\tabularnewline}{\\}

\usepackage{mathtools}
\usepackage{changebar}
\usepackage[noadjust]{cite}
\usepackage{array}
\usepackage[ruled,lined,linesnumbered,longend]{algorithm2e}
\usepackage[font={small}]{caption}
\allowdisplaybreaks[3]

\@ifundefined{showcaptionsetup}{}{%
 \PassOptionsToPackage{caption=false}{subfig}}
\usepackage{subfig}
\makeatother

\begin{document}

\title{Distributed Energy Spectral Efficiency Optimization for Partial/Full
Interference Alignment in Multi-User Multi-Relay Multi-Cell MIMO Systems}

\author{Kent~Tsz~Kan~Cheung, Shaoshi~Yang,~\IEEEmembership{Member,~IEEE},
and Lajos~Hanzo,~\IEEEmembership{Fellow,~IEEE}\thanks{Copyright (c) 2015 IEEE. Personal use of this material is permitted. However, permission to use this material for any other purposes must be obtained from the IEEE by sending a request to pubs-permissions@ieee.org.

This research has been funded by the Industrial Companies who are Members of the Mobile VCE, with additional financial support from the UK Government's Engineering \& Physical Sciences Research Council (EPSRC). The financial support of the Research Councils UK (RCUK) under the India-UK Advanced Technology Center (IU-ATC), of the EU under the auspices of the Concerto project, and of the European Research Council's Senior Research Fellow Grant is also gratefully acknowledged.

The authors are with the School of Electronics and Computer Science, University of Southampton, Southampton, SO17 1BJ, UK (e-mail: \{ktkc106, sy7g09, lh\}@ecs.soton.ac.uk).}\vspace{-10mm}
}

\markboth{Accepted to appear on IEEE Transactions on Signal Processing, Oct. 2015}%
{Shell \MakeLowercase{\textit{et al.}}: Bare Demo of IEEEtran.cls
for Journals}

\maketitle
\begin{abstract}
The energy spectral efficiency maximization~(ESEM) problem of a multi-user,
multi-relay, multi-cell system is considered, where all the network
nodes are equipped with multiple antenna aided transceivers. In order
to deal with the potentially excessive interference originating from
a plethora of geographically distributed transmission sources, a pair
of transmission protocols based on interference alignment~(IA) are
conceived, which may be distributively implemented in the network.
The first, termed the full-IA protocol, avoids all intra-cell interference~(ICI)
and other-cell interference~(OCI) by finding the perfect interference-nulling
receive beamforming matrices~(RxBFMs). The second protocol, termed
as partial-IA, only attempts to null the ICI. Employing the RxBFMs
computed by either of these protocols mathematically decomposes the
channel into a multiplicity of non-interfering multiple-input--single-output~(MISO)
channels, which we term as spatial multiplexing components~(SMCs).
The problem of finding the optimal SMCs as well as their power control
variables for the ESEM problem considered is formally defined and
converted into a convex optimization form with the aid of carefully
selected variable relaxations and transformations. Thus, the optimal
SMCs and power control variables can be distributively computed using
both the classic dual decomposition and subgradient methods. The performance
of both protocols is characterized, and the ESEM algorithm conceived
is compared to a baseline equal power allocation~(EPA) algorithm.
The results indicate that indeed, the ESEM algorithm performs better
than the EPA algorithm in terms of its ESE.
Furthermore, surprisingly the partial-IA protocol outperforms the
full-IA protocol in all cases considered, which may be explained by
the fact that the partial-IA protocol is less restrictive in terms
of the number of available transmit dimensions at the transmitters.
Given the typical cell sizes considered in this paper, the path-loss
sufficiently attenuates the majority of the interference, and thus
the full-IA protocol over-compensates, when trying to avoid all possible
sources of interference. We have observed that, given a sufficiently
high maximum power, the partial-IA protocol achieves an energy spectral
efficiency~(ESE) that is 2.42 times higher than that attained by
the full-IA protocol.\vspace{-5mm}

\end{abstract}

\section{Introduction}

Future wireless cellular networks are required to satisfy ever-increasing
area spectral efficiency~(ASE) demands in the context of densely
packed heterogeneous cells, where both relay nodes~(RNs) and small-cells~\cite{Lopez-Perez2009,Bhat2012}
are employed. However, these changes will result in severe co-channel
interference~(CCI), since future networks will aim for fully exploiting
the precious wireless spectrum by relying on a unity frequency reuse
factor~\cite{El-Ayach2013}. Furthermore, owing to the growing energy
costs, a system's energy efficiency is becoming a major concern~\cite{Han2011}.
\emph{Against this backdrop, in this paper we aim for maximizing the
energy spectral efficiency~(ESE) of the downlink~(DL) of a decode-and-forward~(DF)~\cite{Laneman2004}
relay-aided multiple-input--multiple-output orthogonal frequency division
multiple access~(MIMO-OFDMA) multi-cell network that employs the
technique of interference alignment~(IA).}

IA was first introduced in~\cite{Maddah-Ali2006a,Maddah-Ali2006b,Maddah-Ali2006},
and it was further popularized in~\cite{Maddah-Ali2008,Cadambe2008}.
In~\cite{Cadambe2008}, Cadambe \emph{et al.} described the main
concept of IA and established the attainable degrees of freedom~(DoF),
when employing IA for completely avoiding interference in a network
supporting $K$ user-pairs. The principle of IA is that, instead of
dividing the wireless resources amongst all users~(often termed as
orthogonalization), each user aligns his/her transmissions into a
predetermined subspace, referred to as the interference subspace,
at all the other receivers, so that the remaining subspace at all
receivers becomes free of interference. Thus, the attainable DoFs
in a system supporting $K$ user-pairs is $K/2$ when employing IA,
instead of $1/K$ obtained through orthogonalization~\cite{Cadambe2008}.
This becomes highly favorable, as $K$ increases.

Hence, IA has been advocated as a viable technique of managing the
uplink~(UL) co-channel interference of multi-cell networks~\cite{Suh2008,Gao2014}.
Explicitly, IA is suitable for the UL, since the number of receive
antennas~(RAs) at the basestation~(BS) is typically higher than
the number of transmit antennas~(TAs) at each user equipment~(UE).
Thus, the potentially higher number of signal dimensions available
at the receiver can be exploited for aligning the CCI into a predetermined
interference subspace, so that the BS can receive the transmissions
of its own UEs without CCI. However, this is not feasible in the DL,
since each UE has access to a low number of receive dimensions. This
challenge was successfully tackled by the DL transmission scheme of~\cite{Suh2011},
which relies on specifically designing transmit precoding~(TP) matrices
for reducing the number of transmit dimensions at the BSs, thus facilitating
DL IA at the UEs. In contrast to other IA techniques, such as~\cite{Kim2010,Da2011,Gomadam2011,Rezaee2012,Tang2013},
the technique presented in~\cite{Suh2011} does not require cooperation
among the BSs for exchanging channel state information~(CSI), and
IA is accomplished distributively. Furthermore, this technique facilitates
IA in systems relying on arbitrary antenna configurations with the
aid of frequency- or time-extension, which is capable of substantially
expanding the total number of transmit and receive dimensions in a
multicarrier system such as OFDMA. In~\cite{Yang2013},
the technique of~\cite{Suh2011} was generalized to an arbitrary
number of BSs and UEs, where each of them is equipped with an arbitrary
number of antennas. Furthermore, the authors of~\cite{Yang2013}
employed the semi-orthogonal user selection scheme of Yoo \emph{et
al.}~\cite{Yoo2006} for maximizing the achievable SE. However, relaying
was not considered in~\cite{Yang2013} and each UE was limited to
receiving a single spatial stream.

In this paper, we aim for maximizing the system's attainable ESE,
defined as a counterpart of ASE~\cite{Goldsmith2005}, where the
latter has the units of $\left[\mbox{bits/sec/Hz/m}^{2}\right]$,
while the former is measured in $\left[\mbox{bits/sec/Hz/Joules}\right]$.
This ESE metric has also been utilized in~\cite{Xiong2012,Ng2012a,Devarajan2012,Cheung2013,Cheung2013a,Cheung2014}.
The authors of~\cite{Xiong2012} considered ESE maximization~(ESEM)
of both the UL and the DL of a cellular network, while providing both
the optimal solution method and a lower-complexity heuristic method.
However, the effects of interference were not quantified in the system
model of~\cite{Xiong2012}, since only a single cell was considered.
Additionally, no relaying was employed. In~\cite{Ng2012a}, ESEM
was performed in a multi-cell setting, where the CCI was eliminated
with the aid of BS cooperation~\cite{Gesbert2010} and zero-forcing
beamforming~(ZFBF). However, the authors of~\cite{Ng2012a} have
not considered the benefits of multiple antenna aided nodes or relaying.
As a further advance, the energy-efficiency of a relay aided system
was considered in~\cite{Devarajan2012}, where the objective function~(OF)
of the optimization problem considered was formulated by incorporating
both the spectral efficiency~(SE) and the energy dissipated. Nevertheless,
these two metrics must be appropriately weighted, which is still an
open challenge. Thus, the ESE metric was not formally optimized.

In fact, the maximization of the ESE metric is typically formulated
as a fractional~(in this case, quasi-concave) programming problem~\cite{Dinkelbach1967,Avriel1988,Boyd2004},
which relies on the classic solution methods of the bisection search~\cite{Boyd2004},
and on Dinkelbach's method~\cite{Dinkelbach1967}, as employed in~\cite{Ng2012a,Cheung2013,Cheung2013a}.
However, the bisection search requires the solution of multiple convex
feasibility problems, while Dinkelbach's method requires the solution
of multiple concave subtractive optimization problems. The total number
of algorithmic iterations may become prohibitive in both cases. Hence,
we opt for employing a beneficial method based on the Charnes-Cooper
variable transformation~\cite{Avriel1988,Isheden2012}, allowing
us to solve the ESEM problem by solving a single concave optimization
problem and to demonstrate its benefits to the wireless communications
community.

Let us now elaborate further by classifying the co-channel interference
as intra-cell interference~(ICI) and other-cell interference~(OCI).
In the DL considered, the former describes the interference that a
RN or UE may receive from the BS within its own cell, where multiple
concurrent transmissions are also intended for other RNs or UEs, while
the latter describes the interference originating from sources located
in other cells.

We now provide a concise list of the contributions presented in this
paper.
\begin{itemize}
\item We evaluate the ESEM of IA employed in a realistic
MIMO-OFDMA system involving multiple cells, multiple relays and multiple
users. Although ESEM has been studied intensely in recent years~\cite{Xiong2012,Ng2012a,Devarajan2012},
these contributions typically consider single cells providing coverage
without the assistance of relaying, or do not exploit the benefits
of multiple antenna aided transceivers. Additionally, although IA
was employed recently in~\cite{Kim2010,Sung2010,Alexandropoulos2013,Ronasi2014,Chen2014},
these contributions focus on user-pair networks, rather than on multi-user
cellular networks and the associated challenges of implementing IA
require further research in the latter scenario. More importantly,
previous contributions typically aim for investigating its SE benefits,
while the achievable ESE of using IA-based protocols has not been
explored at all. Green communications has become increasingly important,
but the quantitative benefits of IA have not been documented in the
context of energy-efficient communications. Therefore, in this contribution
we seek to deepen the research community's understanding of IA from
an ESE perspective. Furthermore, a more realistic multi-cell MIMO-OFDMA
relay-aided network is considered in this treatise, where multiple
users are supported by each BS and multiple relays. Therefore, the
system model considered inevitably becomes challenging. As a beneficial
result, the protocols and solutions provided in this paper can be
more readily applied to real network scenarios, when compared to the
existing IA literature, which focuses only on the $K$-user interference
network. In contrast to our previous contributions~\cite{Cheung2013,Cheung2013a,Cheung2014},
this treatise investigates a multiple antenna aided multi-cell system.
Although a multiple antenna assisted system was also studied in our
previous contribution~\cite{Cheung2014}, only a single macrocell
was considered and no IA was employed for avoiding the ICI imposed
by both the simultaneously transmitting BS and RNs.
\item We provide a sophisticated generalization of the
IA protocol considered in~\cite{Suh2011}. Explicitly, in contrast
to~\cite{Suh2011}, the proposed IA protocol accounts for three cells,
for an arbitrary number of users in each cell, for an arbitrary antenna
configuration and for simultaneous direct as well as relay-aided transmissions.
This is accomplished through the careful design of precoding-, transmit-
and receive- beamforming matrices in order to ensure that IA is achieved.
In particular, the number of guaranteed spatial dimensions available
at the BSs, RNs and UEs must be judiciously chosen. Furthermore, we
conceive of two transmission protocols in this work, which may be
implemented distributively at each BS. The first protocol is termed
as full-IA, which invokes IA for avoiding the interference arriving
from all transmitters. This is the intuitive choice, as advocated
by the existing literature~\cite{Suh2008,Suh2011,Yang2013} highlighting
its benefits in terms of achieving the optimal DoF. For example, it
was also employed in~\cite{Suh2011}, but for a simpler system model
having no relays. The second protocol proposed is unlike that of~\cite{Suh2011}
and it is termed as partial-IA, which only aims for avoiding the ICI
using IA, while ignoring the effect of OCI when making scheduling
decisions. The partial-IA protocol therefore reduces
the computational burden of having to estimate the DL CSI of the other-cell
channel matrices at the receivers, albeit this might be expected to
reduce the system's performance due to neglecting the OCI. We
compared the performance of these two protocols and found that, as
a surprise, the reduced-complexity partial-IA protocol is potentially
capable of achieving a higher ESE than the full-IA protocol. Explicitly,
the partial-IA protocol achieves a higher ESE, since more simultaneous
transmissions may be scheduled due to its relaxed constraint on the
number of transmit dimensions available. Furthermore,
in contrast to the protocol proposed in~\cite{Yang2013}, ours is
a two-phase protocol, which is specifically designed for relay-aided
networks and does not limit the number of spatial streams available
to each UE.
\item Employing the beamforming matrices calculated from either the full-IA
or partial-IA protocols results in a list of spatial multiplexing
components~(SMCs)%
\footnote{These SMCs are detailed further in Section~\ref{sec:protocol}%
}, which correspond to the specific data streams that the BSs can choose
to support. Finding the optimal SMCs as well as the optimal power
control variables associated with these optimal SMCs is formally defined
as a network-wide optimization problem. Unlike in
our previous work~\cite{Cheung2013,Cheung2013a,Cheung2014}, we decompose
the network-wide multi-cell optimization problem in order to formulate
a subproblem for each BS using the technique of primal decomposition~\cite{Palomar2006},
thus eliminating the need for the high-overhead backhaul-aided message
passing amongst the BSs. Each of these subproblems is then converted
into a convex form with the aid of various variable relaxations and
transformations, which can then be optimally and distributively solved
using the dual decomposition and subgradient methods of~\cite{Palomar2006}.
\end{itemize}
The organization of this paper is as follows. We introduce our system
model in Section~\ref{sec:model} and describe the proposed transmission
protocols in Section~\ref{sec:protocol}. Subsequently, the ESEM
optimization problem considered is formulated in Section~\ref{sec:optimization},
where the solution method is developed as well. Our numerical results
along with our further discussions are presented in Section~\ref{sec:results}.
Finally, our conclusions are given in Section~\ref{sec:conc} along
with our future research ideas.

\section{System model\label{sec:model}}

\begin{figure}
\begin{centering}
\includegraphics[scale=0.5]{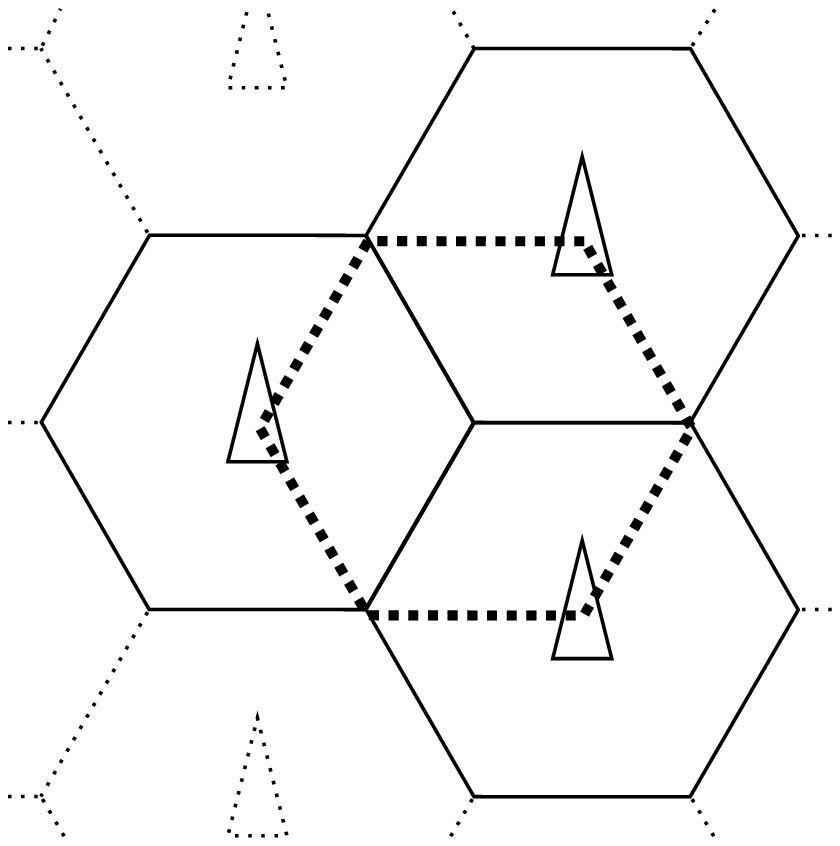}\includegraphics[scale=0.37]{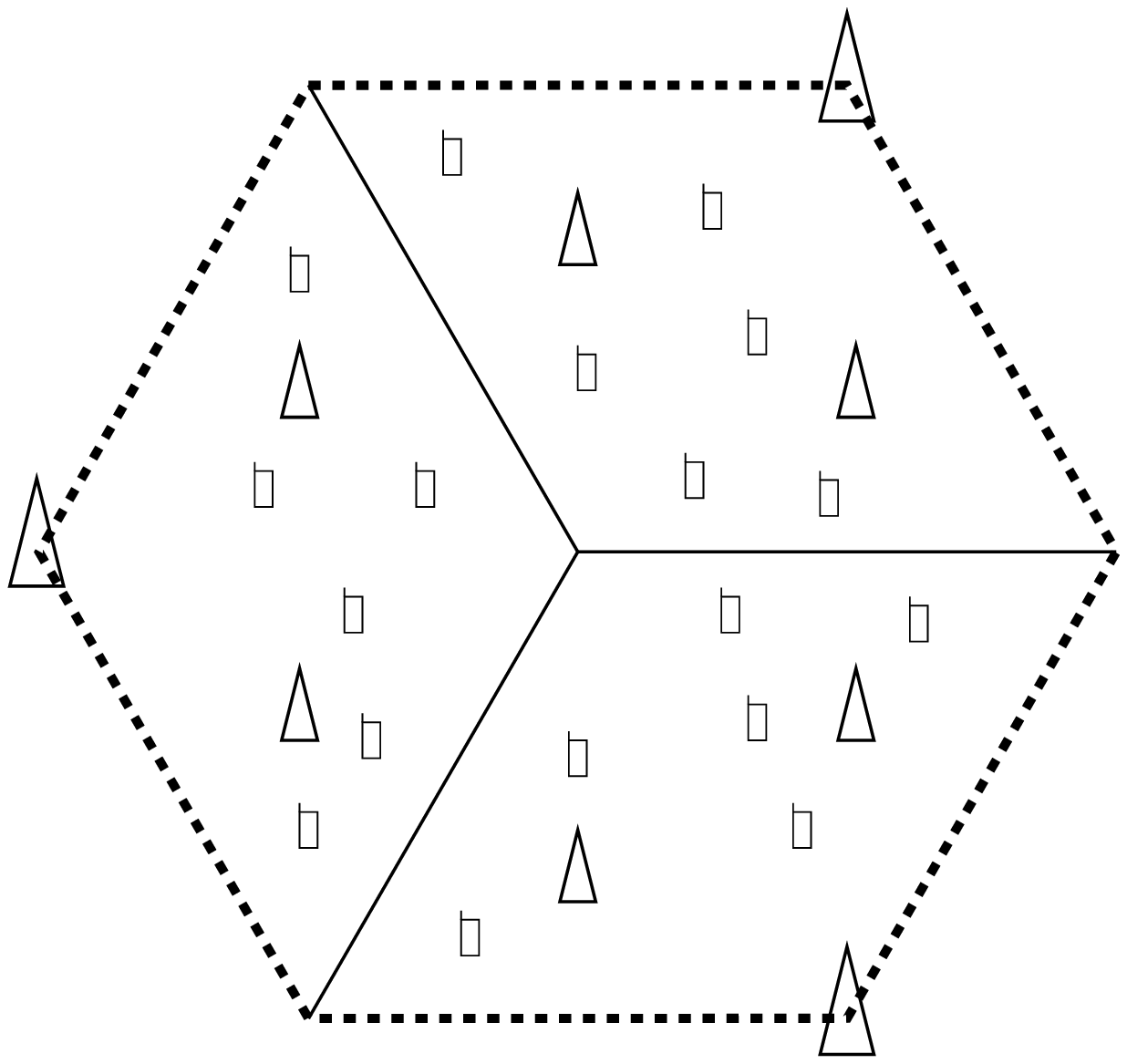}
\par\end{centering}

\caption{A multi-cell system is depicted on the left. Each cell is divided
into three sectors, and one sector from each of the three neighboring
cells are highlighted. This highlighted region is termed an OCI region.
Through the use of directional antennas, it is assumed that the main
source of OCI is caused when the neighboring BSs simultaneously transmit
to a receiver located in its associated OCI region. On the right is
a close-up view of the OCI region, with three BSs at the vertices
of its perimeter. Furthermore, each sector is supported by two RNs
and provides coverage for six UEs in this example.}
\label{fig:model}\vspace{-7mm}
\end{figure}
In this work, a multi-cell DL MIMO-OFDMA network, relying on a radical
unity frequency reuse factor is considered. The
ubiquitous OFDMA technique is employed for avoiding the severe frequency-selective
fading encountered in wideband communication systems. Additionally,
OFDMA allows for transmission symbol extensions in the frequency-domain~\cite{Suh2011},
which are required by the proposed IA-based transmission protocol
described in Section~\ref{sec:protocol}.

As depicted in Fig.~\ref{fig:model}, each macrocell is divided into
three sectors, and it is assumed that the employment of directional
antennas and the non-line-of-sight~(NLOS) path-loss attenuates the
interference power, with the exception of the OCI received from the
first tier of interfering cells and the ICI from the serving BS and
RNs of each macrocell. Therefore, we may focus our attention on the
central region seen at the left of Fig.~\ref{fig:model}, which we
term as an OCI region. Thus, each DL transmission within an OCI region
is subjected to OCI from two macrocells. Furthermore, each $120^{\circ}$-sector
of Fig.~\ref{fig:model} is supported by $M$ RNs, which are located
at a fixed distance from its associated BS and evenly spaced within
the sector, as seen at the right of Fig.~\ref{fig:model}. The ratio
of the BS-RN distance to the cell radius is denoted by $D_{r}$. Additionally,
$K$ UEs are uniformly distributed within each $120^{\circ}$-sector.
The system has access to $L$ OFDMA subcarriers, each characterized
by a wireless bandwidth of $W$ Hertz. The BSs, DF RNs, and UEs are
respectively equipped with $N_{B}$, $N_{R}$ and $N_{U}$ antennas.
It is assumed that all BSs and RNs are synchronized, and that the
transceivers employ complex-valued symbol constellations to convey
their data.

For each subcarrier $l\in\left\{ 1,\cdots,L\right\} $, the complex-valued
channel matrix associated with the wireless link spanning from the
BS of macrocell $n'\in\left\{ 1,2,3\right\} $ to RN $m\in\left\{ 1,\cdots,M\right\} $
belonging to macrocell $n\in\left\{ 1,2,3\right\} $ is denoted by%
\footnote{Superscript indices refer to the transmitter, while subscript indices
refer to the receiver. Additionally, a prime symbol $'$ refers to
a potentially interfering transmission source. %
} $\mathbf{H}_{n,m}^{BR,l,n'}\in\mathbb{C}^{N_{R}\times N_{B}}$. The
channel matrix associated with the link spanning from the BS of macrocell
$n'$ to UE $k\in\left\{ 1,\cdots,K\right\} $ and belonging to macrocell
$n$ on subcarrier $l$ is denoted by $\mathbf{H}_{n,k}^{BU,l,n'}\in\mathbb{C}^{N_{U}\times N_{B}}$.
Furthermore, the channel matrix associated with the link between RN
$m'$ belonging to macrocell $n'$ and UE $k$ belonging to macrocell
$n$ on subcarrier $l$ is denoted by $\mathbf{H}_{n,k}^{RU,l,n',m'}\in\mathbb{C}^{N_{U}\times N_{R}}$.
All channel matrices are assumed to have a full rank, as is often
the case for wireless DL channels. For simplicity, the channel matrices
associated with the same transceivers are combined across subcarriers
to give the block-diagonal channel matrices $\mathbf{H}_{n,m}^{BR,n'}\in\mathbb{C}^{LN_{R}\times LN_{B}}$,
$\mathbf{H}_{n,k}^{BU,n'}\in\mathbb{C}^{LN_{U}\times LN_{B}}$ and
$\mathbf{H}_{n,k}^{RU,n',m'}\in\mathbb{C}^{LN_{U}\times LN_{R}}$,
respectively. For example, we have
\begin{equation}
\mathbf{H}_{n,m}^{BR,n'}\vcentcolon=\left[\begin{array}{ccc}
\mathbf{H}_{n,m}^{BR,1,n'} & \mathbf{0} & \mathbf{0}\\
\mathbf{0} & \ddots & \mathbf{0}\\
\mathbf{0} & \mathbf{0} & \mathbf{H}_{n,m}^{BR,L,n'}
\end{array}\right].
\end{equation}

The channel matrices account for both the small-scale frequency-flat
Rayleigh fading, as well as the large-scale path-loss between the
corresponding transceivers. In this system model, the transceivers
are either stationary or moving sufficiently slowly for ensuring that
the channel matrices can be considered time-invariant for the duration
of a scheduled transmission period. However, the channel matrices
may evolve between each transmission period. Furthermore, it is assumed
that the transceivers' antennas are spaced sufficiently far apart
for ensuring that the associated transmissions experience i.i.d. small-scale
fading, which are drawn from complex i.i.d. normal distributions having
a zero mean and a unit variance. The system uses time-division duplexing~(TDD)
and hence the associated channel reciprocity may be exploited for
predicting the CSI of the slowly varying DL channels from the received
UL signal. Furthermore, by assuming the availability of low-rate error-free
wireless backhaul channels, the CSI associated with the wireless intra-cell
RN-UE links may be fed back to the particular BS in control, so that
it may make the necessary scheduling decisions.

Additionally, each receiver suffers from complex-valued additive white
Gaussian noise~(AWGN) having a power spectral density of $N_{0}$.
Due to both regulatory and safety concerns, the maximum instantaneous
transmission power of each BS and each RN is limited, which are denoted
by $P_{max}^{B}$ and $P_{max}^{R}$, respectively. We stipulate the
idealized simplifying assumption that OFDMA modulation and demodulation
is performed perfectly for all the information symbols.

\section{Transmission protocol design\label{sec:protocol}}

Each BS may convey information to the UEs by either using a direct
BS-UE link, or by utilizing a RN to create a two-hop BS-RN and RN-UE
link, which requires two transmission phases. Thus, each transmission
period is split into two halves. Due to the simultaneous transmissions
from multiple sources, both the level of ICI and OCI in the network
is likely to be detrimental to the achievable ESE. In order to avoid
both types of interference, the technique of IA is employed, which
requires the careful design of both the transmit beamforming matrix~(TxBFMs)
of the BSs and of the RNs, as well as the receive beamforming matrix~(RxBFMs)
of the RNs and of the UEs. As relaying links may be utilized in this
system, the design of these matrices is different for the two transmission
phases. Hence they are described separately in the following. Additionally,
both the full-IA and partial-IA protocols will be described side-by-side.
To elaborate a little further, the full-IA protocol aims for completely
avoiding both the ICI and OCI in both the first and second transmission
phases by employing IA, while the partial-IA protocol only aims for
avoiding the ICI in both transmission phases, thus dispensing with
estimating the OCI channel matrices at each receiver.

\begin{figure*}
\begin{centering}
\includegraphics{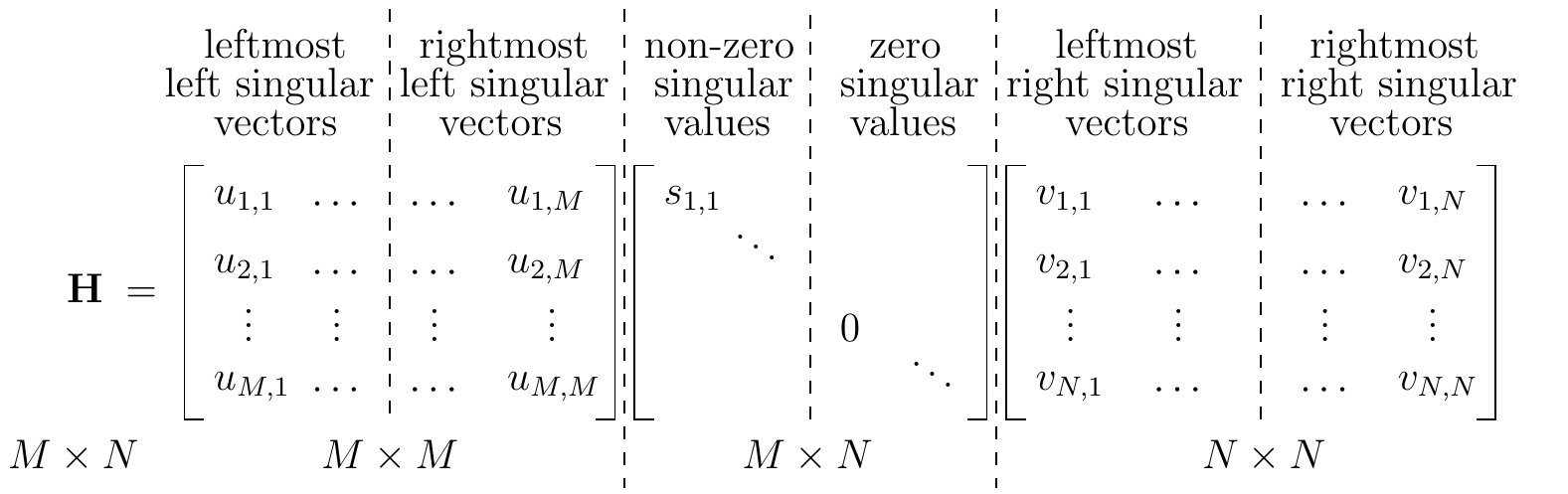}
\par\end{centering}

\caption{The structure of the SVD employed in this paper. The leftmost left
and right singular vectors correspond to the non-zero singular values,
while the rightmost left and right singular vectors correspond to
the zero singular vectors. Therefore, the rightmost left singular
vectors span the left nullspace of $\mathbf{H}$.}
\label{fig:svd}\vspace{-7mm}
\end{figure*}

Furthermore, the proposed schemes crucially rely on the singular value
decomposition~(SVD), where the columns of the left and right singular
matrices are composed of the left and right singular vectors of the
associated matrix. These left and right singular vectors may be further
partitioned into the leftmost and rightmost parts, which correspond
to the non-zero and zero singular values, respectively. This structure
is illustrated in detail in Fig.~\ref{fig:svd}.

\subsection{Beamforming design for the first phase}

In the first phase, only the BSs are transmitting to both the RNs
and the UEs. Therefore, the only source of interference is constituted
by the neighboring BSs associated with the same OCI region, which
may be avoided by carefully designing the TxBFMs at the BSs, as well
as the RxBFMs at the RNs and the UEs in a distributive manner. Initially,
a TP, denoted by $\mathbf{A}^{B,n,T_{1}}\in\mathbb{C}^{LN_{B}\times S^{B,T_{1}}}$,
is randomly-generated for each BS $n$, where $S^{B,T_{1}}$ is the
number of symbols transmitted by each BS during the first phase, which
is accurately defined in Section~\ref{sub:first_phase_scheduling}.
The matrix $\mathbf{A}^{B,n,T_{1}}$ has a full column rank and its
entries are complex-valued. These TPs are invoked for reducing the
number of transmit dimensions for each BS from $LN_{B}$ to $S^{B,T_{1}}$,
thus facilitating IA at the receivers. Furthermore, the columns of
these TP matrices are normalized so that the power assigned to each
transmission remains unaffected. By employing these TPs, the precoded
channel matrices of the first phase are given by
\begin{equation}
\widetilde{\mathbf{H}}_{n,m}^{BR,n',T_{1}}\vcentcolon=\mathbf{H}_{n,m}^{BR,n'}\mathbf{A}^{B,n',T_{1}}\in\mathbb{C}^{LN_{R}\times S^{B,T_{1}}}\label{eq:oci1}
\end{equation}
and
\begin{equation}
\widetilde{\mathbf{H}}_{n,k}^{BU,n',T_{1}}\vcentcolon=\mathbf{H}_{n,k}^{BU,n'}\mathbf{A}^{B,n',T_{1}}\in\mathbb{C}^{LN_{U}\times S^{B,T_{1}}},\label{eq:oci2}
\end{equation}
respectively for the BS-RN and BS-UE links.

We now define $S^{R}$ and $S^{U}$ as the minimum
number of receive dimensions at each RN and each UE, respectively,
which are chosen by the network operator. Furthermore, only the specific
values of $S^{R}$ and $S^{U}$ along with the number of antennas
at each network node and the number of available subcarrier blocks
affect the feasibility of IA, while $M$ and $K$ have no effect.

\subsubsection{Full-IA receiver design}

In order to completely avoid the interference arriving from the neighboring
BSs during the first phase, it is necessary for the precoded OCI channel
matrices given by~(\ref{eq:oci1}) and~(\ref{eq:oci2}), to have
intersecting left nullspaces. Firstly, the precoded OCI channel matrices
may be concatenated for forming the interference matrices, for example
\begin{equation}
\widehat{\mathbf{H}}_{1,m}^{R,T_{1}}\vcentcolon=\left[\widetilde{\mathbf{H}}_{1,m}^{BR,2,T_{1}}\left|\widetilde{\mathbf{H}}_{1,m}^{BR,3,T_{1}}\right.\right]\in\mathbb{C}^{LN_{R}\times2S^{B,T_{1}}}
\end{equation}
for RN $m$ in macrocell $1$, and
\begin{equation}
\widehat{\mathbf{H}}_{2,k}^{U,T_{1}}\vcentcolon=\left[\widetilde{\mathbf{H}}_{2,k}^{BU,1,T_{1}}\left|\widetilde{\mathbf{H}}_{2,k}^{BU,3,T_{1}}\right.\right]\in\mathbb{C}^{LN_{U}\times2S^{B,T_{1}}}
\end{equation}
for UE $k$ in macrocell $2$. These matrices are associated with
a left nullspace of at least $S^{R}$ and $S^{U}$ dimensions if
\begin{equation}
LN_{R}-2S^{B,T_{1}}\geq S^{R}
\end{equation}
and
\begin{equation}
LN_{U}-2S^{B,T_{1}}\geq S^{U},
\end{equation}
respectively. Therefore, to guarantee $S^{R}$ and $S^{U}$ receive
dimensions at the RNs and UEs, respectively, $S^{B,T_{1}}$ is derived
as
\begin{equation}
S^{B,T_{1}}=\left\lfloor \min\left(\frac{LN_{R}-S^{R}}{2},\frac{LN_{U}-S^{U}}{2}\right)\right\rfloor .\label{eq:full_ia_b1}
\end{equation}
The intersecting left nullspace may be found using the SVD on $\widehat{\mathbf{H}}_{n,m}^{R,T_{1}}$
and $\widehat{\mathbf{H}}_{n,k}^{U,T_{1}}$, for RN $m$ and UE $k$
in macrocell $n$, respectively. For example, the SVD of $\widehat{\mathbf{H}}_{n,m}^{R,T_{1}}$
may be written as $\mathbf{U}_{n,m}^{R,T_{1}}\mathbf{S}_{n,m}^{R,T_{1}}\left(\mathbf{V}_{n,m}^{R,T_{1}}\right)^{\mathrm{H}}$,
where $\mathbf{U}_{n,m}^{R,T_{1}}\in\mathbb{C}^{LN_{R}\times LN_{R}}$
is the left singular matrix containing, as its columns, the left singular
vectors of $\widehat{\mathbf{H}}_{n,m}^{R,T_{1}}$, while $\mathbf{S}_{n,m}^{R,T_{1}}\in\mathbb{R}_{+}^{LN_{R}\times2S^{B,T_{1}}}$
is a rectangular diagonal matrix whose diagonal entries are the singular
values of $\widehat{\mathbf{H}}_{n,m}^{R,T_{1}}$ ordered in descending
value, and $\mathbf{V}_{n,m}^{R,T_{1}}\in\mathbb{C}^{2S^{B,T_{1}}\times2S^{B,T_{1}}}$
is the right singular matrix containing, as its columns, the right
singular vectors of $\widehat{\mathbf{H}}_{n,m}^{R,T_{1}}$. The intersecting
left nullspace may then be obtained as the $\left(LN_{R}-2S^{B,T_{1}}\right)$
rightmost columns of $\mathbf{U}_{n,m}^{R,T_{1}}$~(corresponding
to the zero singular values), and this is used as the RxBFM, $\mathbf{R}_{n,m}^{R,T_{1}}$,
for RN $m$ in macrocell $n$. A similar procedure is performed for
obtaining the RxBFM, $\mathbf{R}_{n,k}^{U,T_{1}}$, for UE $k$ in
macrocell $n$ in the first phase, where the $\left(LN_{U}-2S^{B,T_{1}}\right)$
rightmost columns of the corresponding left singular matrix are selected.

To summarize, the cost of implementing the full-IA
protocol in the first transmission phase is the reduction of the number
of available spatial transmission streams at each BS from $LN_{B}$
to $S^{B,T_{1}}$. Thus, if the RNs and UEs require a large number
of spatial streams, the BSs have to substantially reduce the number
of transmitted streams in order to accommodate IA.\textbf{ }However,
it is clear that $S^{B,T_{1}}$ should be higher than $0$ to ensure
that the BSs become capable of transmitting. Following this procedure,
the $\left(3S^{B,T_{1}}-S^{R}\right)$ and $\left(3S^{B,T_{1}}-S^{U}\right)$
total interference signal dimensions received at each RN and at each
UE respectively have each been aligned to $2S^{B,T_{1}}$ dimensions,
leaving $LN_{R}-2S^{B,T_{1}}\geq S^{R}$ and $LN_{U}-2S^{B,T_{1}}\geq S^{U}$
receive signal dimensions free from interference at the RNs and UEs,
respectively. Thus, IA has been successfully employed for reducing
the number of spatial dimensions that the interference signals occupy.

\subsubsection{Partial-IA receiver design}

Using this design philosophy, the OCI encountered during the first
phase is ignored when designing the RxBFMs. However, since there is
no ICI in the first phase since only the BSs are transmitting, there
is no need to reduce the number of transmit dimensions at the BSs.
Therefore, 
\begin{equation}
S^{B,T_{1}}=LN_{B}\label{eq:partial_ia_b1}
\end{equation}
is chosen. Furthermore, the matched filter receiver design is adopted
for maximizing the achievable SE~\cite{Raleigh1998}. In this case,
the SVD is performed on the intra-cell precoded channel matrices,
yielding for example 
\begin{equation}
\widetilde{\mathbf{H}}_{n,m}^{BR,n,T_{1}}=\mathbf{U}_{n,m}^{BR,n,T_{1}}\mathbf{S}_{n,m}^{BR,n,T_{1}}\left(\mathbf{V}_{n,m}^{BR,n,T_{1}}\right)^{\mathrm{H}}
\end{equation}
and 
\begin{equation}
\widetilde{\mathbf{H}}_{n,k}^{BU,n,T_{1}}=\mathbf{U}_{n,k}^{BU,n,T_{1}}\mathbf{S}_{n,k}^{BU,n,T_{1}}\left(\mathbf{V}_{n,k}^{BU,n,T_{1}}\right)^{\mathrm{H}},
\end{equation}
respectively, and the $S^{R}$~(resp. $S^{U}$) leftmost left~(thus
corresponding to the highest singular values) singular vectors are
selected as the RxBFM for the RNs~(resp. UEs) in the first phase.

In summary, IA is not required during the first
transmission phase of the partial-IA protocol, since the only transmitter
within the same cell is the associated BS. Therefore, it is not necessary
for the BSs to reduce the number of transmit dimensions available
to them for the sake of avoiding interference.

\subsubsection{Scheduling and transmitter design\label{sub:first_phase_scheduling}}

Having designed the RxBFMs, the effective DL channel matrices can
be written as
\begin{equation}
\overline{\mathbf{H}}_{n,m}^{BR,n,T_{1}}\vcentcolon=\left(\mathbf{R}_{n,m}^{R,T_{1}}\right)^{\mathrm{H}}\widetilde{\mathbf{H}}_{n,m}^{BR,n,T_{1}}
\end{equation}
or
\begin{equation}
\overline{\mathbf{H}}_{n,k}^{BU,n,T_{1}}\vcentcolon=\left(\mathbf{R}_{n,k}^{U,T_{1}}\right)^{\mathrm{H}}\widetilde{\mathbf{H}}_{n,k}^{BU,n,T_{1}}
\end{equation}
for RN $m$ and UE $k$ in macrocell $n$, respectively. We term the
rows of these matrices as the SMCs of the associated transceivers,
since each SMC corresponds to a distinct \emph{virtual} multiple-input--single-output~(MISO)
channel between the associated transmitter as well as receiver, and
then multiple MISOs can be multiplexed for composing a MIMO channel.
A set of SMCs is generated for each of the two transmission phases,
and each BS then distributively groups these SMCs according to the
semi-orthogonal user selection algorithm%
\footnote{This selection method aims for reducing the power loss imposed by
the channel inversion operation of the ZFBF matrix~\cite{Yoo2006,Cheung2014}.%
}, as described in~\cite{Yoo2006,UlHassan2009}, given a semi-orthogonality
parameter $\alpha$. For the first transmission phase, up to $\min\left(S^{B,T_{1}},KLN_{U}+MLN_{R}\right)$
SMCs may be served simultaneously by each BS, while avoiding ICI.
The set of groupings available for BS $n$ is denoted by%
\footnote{N.B. Each group additionally contains the SMCs selected for the second
phase, as it will be discussed in Section~\ref{sub:second_phase_scheduling}.%
} $\mathcal{G}^{n}$. The SMCs belonging to group $j$, which are denoted
by $\mathcal{E}^{n,j}$, are then the rows of the effective scheduled
DL matrix, denoted by $\underline{\mathbf{H}}^{B,n,j,T_{1}}$ for
macrocell $n$. In order to avoid ICI between these selected SMCs
of group $j$, macrocell $n$ applies the ZFBF matrix, given in~(\ref{eq:zfbf_ph1}),
by $\mathbf{T}^{B,n,j,T_{1}}$ as the right channel inverse before
using its TP, $\mathbf{A}^{B,n}$, where $\left(\mathbf{W}^{B,n,j,T_{1}}\right)^{\frac{1}{2}}$
is a real-valued diagonal matrix, which normalizes the columns of
$\mathbf{T}^{B,n,j,T_{1}}$ for ensuring that the power assigned to
each transmission remains unaffected.
\begin{figure*}
\begin{equation}
\mathbf{T}^{B,n,j,T_{1}}=\left(\underline{\mathbf{H}}^{B,n,j,T_{1}}\right)^{\mathrm{H}}\left[\underline{\mathbf{H}}^{B,n,j,T_{1}}\left(\underline{\mathbf{H}}^{B,n,j,T_{1}}\right)^{\mathrm{H}}\right]^{-1}\left(\mathbf{W}^{B,n,j,T_{1}}\right)^{\frac{1}{2}}\label{eq:zfbf_ph1}
\end{equation}
\hrulefill

\vspace{-5mm}
\end{figure*}

The effective end-to-end channel power gains are then given by the
squares of the diagonal entries in $\left(\mathbf{W}^{B,n,j,T_{1}}\right)^{\frac{1}{2}}$.
For SMC $e_{1}$ in group $j$ of macrocell $n$ corresponding to
a direct first phase BS-UE link, the effective channel power gain
is denoted by $w_{n,e_{1}}^{BU,n,j,T_{1}}$, while the effective channel
power gain of the OCI link, originating from macrocell $n'$ serving
SMC group $j'$ to UE $k$ in macrocell $n$, is obtained from the
element of
\begin{equation}
\left|\left(\mathbf{R}_{n,k}^{U,T_{1}}\right)^{\mathrm{H}}\widetilde{\mathbf{H}}_{n,k}^{BU,n',T_{1}}\mathbf{T}^{B,n',j',T_{1}}\right|^{2}
\end{equation}
corresponding to SMC $e_{1}$ at UE $k$ of macrocell $n$, and is
denoted by $w_{n,e_{1}}^{BU,n',j',T_{1}}$. In the case of the full-IA
protocol, all OCI is avoided, thus $w_{n,e_{1}}^{BU,n',j',T_{1}}=0$,
$\forall n'\neq n$. The effective channel power gains for the BS-to-RN
links, corresponding to SMC-pair%
\footnote{Relaying links contain both a SMC for the BS-RN link and a SMC for
the RN-UE link.%
} $e$, may be similarly obtained and are denoted by $w_{n,e}^{BR,n,j,T_{1}}$,
whereas an OCI link is denoted by $w_{n,e}^{BR,n',j',T_{1}}$.

\subsection{Beamforming design for the second phase}

During the second phase, both the BSs and the RNs may transmit. Therefore,
in a similar fashion to the first phase, the BS in cell $n$ adopts
the precoding matrix $\mathbf{A}^{B,n,T_{2}}\in\mathbb{C}^{LN_{B}\times S^{B,T_{2}}}$,
while RN $m$ in cell $n$ adopts the precoding matrix $\mathbf{A}^{R,n,m,T_{2}}\in\mathbb{C}^{LN_{R}\times S^{R}}$,
which are again complex-valued matrices having a full column-rank.
Additionally, the columns of these TP matrices are normalized. Due
to the additional interference imposed by the transmissions of the
RNs, it is necessary to reduce the number of transmit dimensions at
the BSs even further in order to facilitate IA at the DL receivers.
Additionally, note that each TP matrix used at the RNs consist of
$S^{R}$ columns, since the information received by each RN during
the first phase must be conveyed to the intended UE. The precoded
channel matrices used during the second phase are thus given by~(note
that the transmitter indices are $n'$ and $m'$, since these may
be inter-cell channel matrices) 
\begin{equation}
\widetilde{\mathbf{H}}_{n,k}^{RU,n',m',T_{2}}\vcentcolon=\mathbf{H}_{n,k}^{RU,n',m'}\mathbf{A}^{R,n',m',T_{2}}\in\mathbb{C}^{LN_{U}\times S^{R}}
\end{equation}
and
\begin{equation}
\widetilde{\mathbf{H}}_{n,k}^{BU,n',T_{2}}\vcentcolon=\mathbf{H}_{n,k}^{BU,n'}\mathbf{A}^{B,n',T_{2}}\in\mathbb{C}^{LN_{U}\times S^{B,T_{2}}}.
\end{equation}

\subsubsection{Full-IA receiver design}

The receiver design used during the second phase depends on whether
the BS or a RN is selected to serve each UE within the same macrocell.
Each of the $(1+M)$ possible transmitters may be examined for the
sake of finding the most beneficial choice. For example, assuming
that BS $1$ transmits to UE $k$ during the second phase, the OCI
and ICI channel matrices may be concatenated to form~(\ref{eq:int_matB_ph2}).
However, when assuming for example, that RN $1$ of macrocell $n$
transmits to UE $k$, the combined interference matrix is defined
by~(\ref{eq:int_matR_ph2}).
\begin{figure*}
\begin{eqnarray}
\widehat{\mathbf{H}}_{1,k}^{BU,1,T_{2}}\vcentcolon & = & \left[\widetilde{\mathbf{H}}_{1,k}^{BU,2,T_{2}}\right|\widetilde{\mathbf{H}}_{1,k}^{BU,3,T_{2}}\left|\widetilde{\mathbf{H}}_{1,k}^{RU,1,1,T_{2}}\right|\cdots\left|\widetilde{\mathbf{H}}_{1,k}^{RU,1,M,T_{2}}\right.\nonumber \\
 &  & \left|\widetilde{\mathbf{H}}_{1,k}^{RU,2,1,T_{2}}\right|\cdots\left|\widetilde{\mathbf{H}}_{1,k}^{RU,2,M,T_{2}}\right.\left|\widetilde{\mathbf{H}}_{1,k}^{RU,3,1,T_{2}}\right|\cdots\left|\widetilde{\mathbf{H}}_{1,k}^{RU,3,M,T_{2}}\right]\in\mathbb{C}^{LN_{U}\times\left(2S^{B,T_{2}}+3MS^{R}\right)}\label{eq:int_matB_ph2}
\end{eqnarray}
\begin{eqnarray}
\widehat{\mathbf{H}}_{n,k}^{RU,n,1,T_{2}}\vcentcolon & = & \left[\widetilde{\mathbf{H}}_{n,k}^{BU,1,T_{2}}\right|\widetilde{\mathbf{H}}_{n,k}^{BU,2,T_{2}}\left|\widetilde{\mathbf{H}}_{n,k}^{BU,3,T_{2}}\right.\left|\widetilde{\mathbf{H}}_{n,k}^{RU,1,2,T_{2}}\right|\cdots\left|\widetilde{\mathbf{H}}_{n,k}^{RU,1,M,T_{2}}\right.\nonumber \\
 &  & \left|\widetilde{\mathbf{H}}_{n,k}^{RU,2,1,T_{2}}\right|\cdots\left|\widetilde{\mathbf{H}}_{n,k}^{RU,2,M,T_{2}}\right.\left|\widetilde{\mathbf{H}}_{n,k}^{RU,3,1,T_{2}}\right|\cdots\left|\widetilde{\mathbf{H}}_{n,k}^{RU,3,M,T_{2}}\right]\in\mathbb{C}^{LN_{U}\times\left[3S^{B,T_{2}}+\left(3M-1\right)S^{R}\right]}\label{eq:int_matR_ph2}
\end{eqnarray}
\hrulefill
\end{figure*}
 Therefore, in order to guarantee having $S^{U}$ receive dimensions
at each UE, we have
\begin{eqnarray}
S^{B,T_{2}} & = & \left\lfloor \min\left(\frac{LN_{U}-S^{U}-3MS^{R}}{2},\right.\right.\nonumber \\
 &  & \left.\left.\frac{LN_{U}-S^{U}-\left(3M-1\right)S^{R}}{3}\right)\right\rfloor .\label{eq:full_ia_b2}
\end{eqnarray}
In both cases described above, the SVD may again be employed for finding
the intersecting left nullspace of the precoded interference matrix.
The RxBFM, $\mathbf{R}_{n,k}^{U,T_{2}}$, at UE $k$ in macrocell
$n$ used during the second phase is then given by the rightmost~(thus
corresponding to its zero singular values) $LN_{U}-\left(2S^{B,T_{2}}+3MS^{R}\right)$
number of columns in the left singular matrix of $\widehat{\mathbf{H}}_{1,k}^{BU,1,T_{2}}$,
when the BS is the activated transmitter. By contrast, when assuming
that RN $1$ is the activated transmitter, the rightmost $\min\left(S^{R},LN_{U}-\left[3S^{B,T_{2}}+\left(3M-1\right)S^{R}\right]\right)$
number of columns in the ordered left singular matrix of $\widehat{\mathbf{H}}_{n,k}^{RU,n,1,T_{2}}$
specify the RxBFM matrix.

In conclusion, the BSs once again have to reduce
the number of spatial transmission streams available to them in order
to facilitate IA. In this case, their number is reduced from $LN_{B}$
to $S^{B,T_{2}}$. Additionally, each RN reduces the number of streams
available for them to transmit from $LN_{R}$ to $S_{R}$. On one
hand, when the BS is selected as the active transmitter for a particular
UE using the full-IA protocol, a total of $\left(3S^{B,T_{2}}+3MS^{R}-S^{U}\right)$
interference signal dimensions are aligned to $\left(2S^{B,T_{2}}+3MS^{R}\right)$
signal dimensions, leaving $LN_{U}-\left(2S^{B,T_{2}}+3MS^{R}\right)\geq S^{U}$
signal dimensions free from interference. Thus, IA has been successfully
employed. On the other hand, when a RN is selected as the activated
transmitter for a particular UE, there is a total of $\left(3S^{B,T_{2}}+3MS^{R}-S^{U}\right)$
interference signal dimensions, which are aligned to $\left(3S^{B,T_{2}}+3MS^{R}-S^{R}\right)$
signal dimensions. Therefore, IA is only feasible at the UEs if we
have $S^{R}>S^{U}$. The constraint given by
\begin{equation}
LN_{U}-\left(3S^{B,T_{2}}+3MS^{R}\right)-S^{R}>S^{R}\label{eq:sr_constraint}
\end{equation}
is additionally enforced in the full-IA protocol, so that the CCI
can still be nulled when $S^{R}\leq S^{U}$ and a RN is selected as
the active transmitter. However, IA is not employed in this case.

\subsubsection{Partial-IA receiver design}

Although the effects of OCI are ignored when using this protocol,
the ICI must be avoided. Thus, the interference matrix, assuming for
example that the BS is the selected transmitter for UE $k$ in macrocell
$1$, is then given by
\begin{equation}
\widehat{\mathbf{H}}_{1,k}^{BU,1,T_{2}}\vcentcolon=\left[\widetilde{\mathbf{H}}_{1,k}^{RU,1,1,T_{2}}\right|\cdots\left|\widetilde{\mathbf{H}}_{1,k}^{RU,1,M,T_{2}}\right]\in\mathbb{C}^{LN_{U}\times MS^{R}}.
\end{equation}
By contrast, if RN $1$ of macrocell $n$ is selected as the transmitter
for UE $k$, then the interference matrix is given by
\begin{eqnarray}
\widehat{\mathbf{H}}_{n,k}^{RU,n,1,T_{2}} & \vcentcolon= & \left[\widetilde{\mathbf{H}}_{n,k}^{BU,1,T_{2}}\right|\left.\widetilde{\mathbf{H}}_{n,k}^{RU,n,2,T_{2}}\right|\cdots\left|\widetilde{\mathbf{H}}_{n,k}^{RU,n,M,T_{2}}\right]\nonumber \\
 &  & \in\mathbb{C}^{LN_{U}\times\left[S^{B,T_{2}}+\left(M-1\right)S^{R}\right]},
\end{eqnarray}
which implies that
\begin{equation}
S^{B,T_{2}}=LN_{U}-S^{U}-\left(M-1\right)S^{R}\label{eq:partial_ia_b2}
\end{equation}
is satisfied for ensuring that the UEs are capable of finding approximate
RxBFMs, which completely null the ICI. 

Thus, UE $k$ may employ the $LN_{U}-MS^{R}$ number of rightmost
left singular columns in $\widehat{\mathbf{H}}_{1,k}^{BU,1,T_{2}}$
as its RxBFM, when the BS is the activated transmitter. By contrast,
assuming that RN $1$ is the activated transmitter, the $\min\left(S^{R},LN_{U}-\left[S^{B,T_{2}}+\left(M-1\right)S^{R}\right]\right)$
number of rightmost left singular columns in $\widehat{\mathbf{H}}_{n,k}^{RU,n,1,T_{2}}$
specify the RxBFM.

To summarize, the BSs reduce the number of spatial
streams available to them from $LN_{B}$ to $S^{B,T_{2}}$, while
the RNs reduce the number of their spatial streams from $LN_{R}$
to $S^{R}$. On one hand, when the BS is selected as the active transmitter
for the partial-IA protocol, a total of $\left(S^{B,T_{2}}+MS^{R}-S^{U}\right)$
interference signal dimensions are aligned to $MS^{R}$ signal dimensions,
leaving $LN_{U}-MS^{R}\geq S^{U}$ signal dimensions free from interference.
Thus, IA has been successfully employed. On the other hand, when a
RN is selected as the activated transmitter, there are a total of
$\left(S^{B,T_{2}}+MS^{R}-S^{U}\right)$ interference signal dimensions,
which are aligned to $\left(S^{B,T_{2}}+MS^{R}-S^{R}\right)$ signal
dimensions. Therefore, IA is only feasible for $S^{R}>S^{U}$. However,
the aforementioned RxBFMs are still capable of nulling the CCI, when
a RN is selected as the active transmitter in the partial-IA protocol
and we have $S^{R}\leq S^{U}$. But in this case the constraint given
by~(\ref{eq:sr_constraint}) is not required, since it is already
satisfied by~(\ref{eq:partial_ia_b2}).

\subsubsection{Scheduling and transmitter design\label{sub:second_phase_scheduling}}

In a similar fashion to the first phase, the effective DL channel
matrices are given by
\begin{equation}
\overline{\mathbf{H}}_{n,k}^{RU,n,m,T_{2}}\vcentcolon=\left(\mathbf{R}_{n,k}^{U,T_{2}}\right)^{\mathrm{H}}\widetilde{\mathbf{H}}_{n,k}^{RU,n,m,T_{2}}
\end{equation}
and
\begin{equation}
\overline{\mathbf{H}}_{n,k}^{BU,n,T_{2}}\vcentcolon=\left(\mathbf{R}_{n,k}^{U,T_{2}}\right)^{\mathrm{H}}\widetilde{\mathbf{H}}_{n,k}^{BU,n,T_{2}},
\end{equation}
when the BS or RN $m$ is activated as the transmitter for UE $k$
belonging to macrocell $n$, respectively. The rows of the DL TxBFMs
corresponding to each transmitter form the SMCs for that transmitter,
and they may be grouped at each BS according to the semi-orthogonal
user selection algorithm described above. Furthermore, in the second
phase, each BS can select up to $\min\left(S^{B,T_{2}},KLN_{U}\right)$
number of SMCs to serve simultaneously while avoiding ICI, whereas
each RN may select $\min\left(S^{R},KLN_{U}\right)$ number of SMCs.
At BS $n$~(or RN $m$ of macrocell $n$), the selected SMCs of group
$j$ form the rows of its effective scheduled DL matrix, denoted by
$\underline{\mathbf{H}}^{B,n,j,T_{2}}$~(or $\underline{\mathbf{H}}^{R,n,m,j,T_{2}}$).
The ZFBF matrix employed by BS $n$ or by RN $m$ of macrocell $n$
in the second phase is then given by the right inverse~(\ref{eq:zfbfB_ph2})
or ~(\ref{eq:zfbfR_ph2}), respectively, where the real-valued diagonal
matrices of $\left(\mathbf{W}^{B,n,j,T_{2}}\right)^{\frac{1}{2}}$
and $\left(\mathbf{W}^{R,n,m,j,T_{2}}\right)^{\frac{1}{2}}$ are required
for normalizing the columns of $\mathbf{T}^{B,n,j,T_{2}}$ and $\mathbf{T}^{R,n,m,j,T_{2}}$,
respectively.
\begin{figure*}
\begin{equation}
\mathbf{T}^{B,n,j,T_{2}}=\left(\underline{\mathbf{H}}^{B,n,j,T_{2}}\right)^{\mathrm{H}}\left[\underline{\mathbf{H}}^{B,n,j,T_{2}}\left(\underline{\mathbf{H}}^{B,n,j,T_{2}}\right)^{\mathrm{H}}\right]^{-1}\left(\mathbf{W}^{B,n,j,T_{2}}\right)^{\frac{1}{2}}\label{eq:zfbfB_ph2}
\end{equation}
\begin{equation}
\mathbf{T}^{R,n,m,j,T_{2}}=\left(\underline{\mathbf{H}}^{R,n,m,j,T_{2}}\right)^{\mathrm{H}}\left[\underline{\mathbf{H}}^{R,n,m,j,T_{2}}\left(\underline{\mathbf{H}}^{R,n,m,j,T_{2}}\right)^{\mathrm{H}}\right]^{-1}\left(\mathbf{W}^{R,n,m,j,T_{2}}\right)^{\frac{1}{2}}\label{eq:zfbfR_ph2}
\end{equation}

\hrulefill\vspace{-5mm}
\end{figure*}

The effective channel power gains in the second phase are thus given
by the squares of the diagonal entries in $\left(\mathbf{W}^{B,n,j,T_{2}}\right)^{\frac{1}{2}}$
and $\left(\mathbf{W}^{R,n,m,j,T_{2}}\right)^{\frac{1}{2}}$. The
effective channel power gain of a BS-UE SMC $e_{2}$ of group $j$
associated with macrocell $n$ and UE $k$ is denoted by $w_{n,e_{2}}^{BU,n,j,T_{2}}$,
while the RN-UE effective channel power gain of SMC-pair $e$ associated
with RN $m$ of macrocell $n$ and UE $k$ may be denoted by $w_{n,e}^{RU,n,m,j,T_{2}}$.
Similar to the first phase, the effective channel power gain of the
OCI link originating from the BS of macrocell $n'$ serving group
$j'$ to UE $k$ in macrocell $n$, is obtained from the specific
element of
\begin{equation}
\left|\left(\mathbf{R}_{n,k}^{U,T_{2}}\right)^{\mathrm{H}}\widetilde{\mathbf{H}}_{n,k}^{BU,n',T_{2}}\mathbf{T}^{B,n',j',T_{2}}\right|^{2}
\end{equation}
corresponding to SMC $e_{2}$ at UE $k$ of macrocell $n$, which
is denoted by $w_{n,e_{2}}^{BU,n',j',T_{2}}$. On the other hand,
the effective channel power gain of the OCI link, originating from
RN $m'$ of macrocell $n'$ serving group $j'$ to UE $k$ of macrocell
$n$, is obtained from the element of
\begin{equation}
\left|\left(\mathbf{R}_{n,k}^{U,T_{2}}\right)^{\mathrm{H}}\widetilde{\mathbf{H}}_{n,k}^{RU,n',m',T_{2}}\mathbf{T}^{R,n',m',j,T_{2}}\right|^{2}
\end{equation}
corresponding to SMC $e$ at UE $k$ of macrocell $n$, and is denoted
by $w_{n,e}^{RU,n',m',j',T_{2}}$. In the case of the full-IA protocol,
all OCI is avoided, thus we have $w_{n,e_{2}}^{BU,n',j',T_{2}}=w_{n,e}^{RU,n',m',j',T_{2}}=0$,
$\forall n'\neq n$.

\subsection{Achievable spectral efficiency and energy efficiency}

Since we have mathematically decomposed the MIMO
channels into effective SISO channels, we may directly employ the
Shannon capacity bound for characterizing the achievable ESE performance,
rather than relying on bounds derived for MIMO channels~\cite{Blum2003}.
We begin by defining the signal-to-interference-plus-noise-ratio~(SINR)
of the direct link SMCs belonging to group $j$ and intended for UE
$k$ of macrocell $n$ during the first and the second phase as
\begin{equation}
\Gamma_{n,e_{1}}^{BU,n,j,T_{1}}\left(\mathcal{P},\mathcal{S}\right)=\frac{w_{n,e_{1}}^{BU,n,j,T_{1}}P_{n,e_{1}}^{B,n,j,T_{1}}}{\Delta\gamma\left(N_{0}LW+I_{n,e_{1}}^{U,T_{1}}\right)}
\end{equation}
and
\begin{equation}
\Gamma_{n,e_{2}}^{BU,n,j,T_{2}}\left(\mathcal{P},\mathcal{S}\right)=\frac{w_{n,e_{2}}^{BU,n,j,T_{2}}P_{n,e_{2}}^{B,n,j,T_{2}}}{\Delta\gamma\left(N_{0}LW+I_{n,e_{2}}^{U,T_{2}}\right)},
\end{equation}
respectively, where the total received OCI in the first and second
phase has been denoted by~(\ref{eq:direct_int_ph1}) and~(\ref{eq:direct_int_ph2}),
respectively, where $\mathcal{M}\left(e\right)$ is a function of
$e$, representing the RN index~(similar to $m$ used before) associated
with the SMC-pair $e$. For simplicity%
\footnote{If the level of interference is strong enough, then more sophisticated
methods, such as multiuser detection, may be employed.%
}, the interference that was not avoided using IA is treated as noise.
\begin{figure*}
\begin{equation}
I_{n,e_{1}}^{U,T_{1}}\left(\mathcal{P},\mathcal{S}\right)=\sum_{\substack{n'=1\\
n'\neq n
}
}\sum_{j'\in\mathcal{G}^{n'}}s^{n',j'}w_{n,e_{1}}^{BU,n',j',T_{1}}\left[\sum_{e'_{1}\in\mathcal{E}^{n',j'}}P_{n',e'_{1}}^{B,n',j',T_{1}}+\sum_{e'\in\mathcal{E}^{n',j'}}P_{n',e'}^{B,n',j',T_{1}}\right]\label{eq:direct_int_ph1}
\end{equation}
\begin{equation}
I_{n,e_{2}}^{U,T_{2}}\left(\mathcal{P},\mathcal{S}\right)=\sum_{\substack{n'=1\\
n'\neq n
}
}\sum_{j'\in\mathcal{G}^{n'}}s^{n',j'}\left[w_{n,e_{2}}^{BU,n',j',T_{2}}\sum_{e'_{2}\in\mathcal{E}^{n',j'}}P_{n',e_{2}'}^{B,n',j',T_{2}}+\sum_{e'\in\mathcal{E}^{n',j'}}w_{n,e_{2}}^{RU,n',\mathcal{M}\left(e'\right),j',T_{2}}P_{n,e'}^{R,n',\mathcal{M}\left(e'\right),j',T_{2}}\right]\label{eq:direct_int_ph2}
\end{equation}

\hrulefill\vspace{-5mm}
\end{figure*}
The set $\mathcal{P}$ contains the power control variables denoted
by $P_{n,e_{1}}^{B,n,j,T_{1}}$, $P_{n,e}^{B,n,j,T_{1}}$, $P_{n,e_{2}}^{B,n,j,T_{2}}$,
and $P_{n,e}^{R,n,m,j,T_{2}}$, $\forall n,e_{1},e_{2},e$. On the
other hand, the set $\mathcal{S}$ contains the group selection indicator
variables, $s^{n,j}$, $\forall n,j$, where $s^{n,j}=1$, when the
SMC group $j$ has been selected for macrocell $n$, and $s^{n,j}=0$
otherwise. The total noise power across all subcarriers is given by
$N_{0}LW$, while $\Delta\gamma$ is the signal to noise ratio~(SNR)
difference between the SNR at the discrete-input\textendash{}continuous-output
memoryless channel~(DCMC) capacity and the actual SNR required by
the specific modulation and coding schemes of the practical physical
layer transceivers employed~\cite{Hanzo2009}.

The SINR of the BS-RN SMC $e$ belonging to group $j$ of macrocell
$n$ and intended for RN $m$ may be expressed as
\begin{equation}
\Gamma_{n,e}^{BR,n,j,T_{1}}\left(\mathcal{P},\mathcal{S}\right)=\frac{w_{n,e}^{BR,n,j,T_{1}}P_{n,e}^{B,n,j,T_{1}}}{\Delta\gamma\left(N_{0}LW+I_{n,e}^{R,T_{1}}\right)},
\end{equation}
while the SINR of the corresponding RN-UE link may be formulated as
\begin{equation}
\Gamma_{n,e}^{RU,n,m,j,T_{2}}\left(\mathcal{P},\mathcal{S}\right)=\frac{w_{n,e}^{RU,n,m,j,T_{2}}P_{n,e}^{R,n,m,j,T_{2}}}{\Delta\gamma\left(N_{0}LW+I_{n,e}^{U,T_{2}}\right)},
\end{equation}
where the total received OCI of the BS-RN and RN-UE links are given
by~(\ref{eq:relayed_int_ph1}) and~(\ref{eq:relayed_int_ph2}),
respectively.
\begin{figure*}
\begin{equation}
I_{n,e}^{R,T_{1}}\left(\mathcal{P},\mathcal{S}\right)=\sum_{\substack{n'=1\\
n'\neq n
}
}\sum_{j'\in\mathcal{G}^{n'}}s^{n',j'}w_{n,e}^{BR,n',j',T_{1}}\left[\sum_{e_{1}'\in\mathcal{E}^{n',j'}}P_{n',e_{1}'}^{B,n',j',T_{1}}+\sum_{e'\in\mathcal{E}^{n',j'}}P_{n',e'}^{B,n',j',T_{1}}\right]\label{eq:relayed_int_ph1}
\end{equation}
\begin{equation}
I_{n,e}^{U,T_{2}}\left(\mathcal{P},\mathcal{S}\right)=\sum_{\substack{n'=1\\
n'\neq n
}
}\sum_{j'\in\mathcal{G}^{n'}}s^{n',j'}\left[w_{n,e}^{BU,n',j',T_{2}}\sum_{e_{2}'\in\mathcal{E}^{n',j'}}P_{n',e'_{2}}^{B,n',j',T_{2}}+\sum_{e'\in\mathcal{E}^{n',j'}}w_{n,e}^{RU,n',\mathcal{M}\left(e'\right),j',T_{2}}P_{n',e'}^{R,n',\mathcal{M}\left(e'\right),j',T_{2}}\right]\label{eq:relayed_int_ph2}
\end{equation}

\hrulefill\vspace{-3mm}
\end{figure*}

The achievable SE of the direct first and second phase transmissions
can be respectively written as
\begin{equation}
C_{n,e_{1}}^{BU,n,j,T_{1}}\left(\mathcal{P},\mathcal{S}\right)=\frac{1}{2}\log_{2}\left(1+\Gamma_{n,e_{1}}^{BU,n,j,T_{1}}\right)\label{eq:c_direct1}
\end{equation}
and
\begin{equation}
C_{n,e_{2}}^{BU,n,j,T_{2}}\left(\mathcal{P},\mathcal{S}\right)=\frac{1}{2}\log_{2}\left(1+\Gamma_{n,e_{2}}^{BU,n,j,T_{2}}\right),\label{eq:c_direct2}
\end{equation}
where the pre-log factor of $\frac{1}{2}$ accounts for the fact that
the transmission period has been split into two phases. When using
the DF protocol, the achievable SE of the relaying link is limited
by the weaker of the BS-RN and RN-UE links~\cite{Laneman2004}, which
is given by
\begin{eqnarray}
C_{n,e}^{BRU,n,m,j}\left(\mathcal{P},\mathcal{S}\right) & = & \min\left[\frac{1}{2}\log_{2}\left(1+\Gamma_{n,e}^{BR,n,j,T_{1}}\right),\right.\nonumber \\
 &  & \left.\frac{1}{2}\log_{2}\left(1+\Gamma_{n,e}^{RU,n,m,j,T_{2}}\right)\right].\nonumber \\
\label{eq:c_relayed}
\end{eqnarray}
Thus the total achievable SE of macrocell $n$ is given by~(\ref{eq:CnT}).
\begin{figure*}
\begin{equation}
C_{T}^{n}\left(\mathcal{P},\mathcal{S}\right)=\sum_{j\in\mathcal{G}^{n}}s^{n,j}\left[\sum_{e_{1}\in\mathcal{E}^{n,j}}C_{n,e_{1}}^{BU,n,j,T_{1}}+\sum_{e_{2}\in\mathcal{E}^{n,j}}C_{n,e_{2}}^{BU,n,j,T_{2}}+\sum_{e\in\mathcal{E}^{n,j}}C_{n,e}^{BRU,n,\mathcal{M}\left(e\right),j}\right]\label{eq:CnT}
\end{equation}
\begin{eqnarray}
P_{T}^{n}\left(\mathcal{P},\mathcal{S}\right) & = & \left(P_{C}^{B}+MP_{C}^{R}\right)\nonumber \\
 &  & +\frac{1}{2}\sum_{j\in\mathcal{G}^{n}}s^{n,j}\left[\xi^{B}\left(\sum_{e_{1}\in\mathcal{E}^{n,j}}P_{n,e_{1}}^{B,n,j,T_{1}}+\sum_{e_{2}\in\mathcal{E}^{n,j}}P_{n,e_{2}}^{B,n,j,T_{2}}\right)+\sum_{e\in\mathcal{E}^{n,j}}\left(\xi^{B}P_{n,e}^{B,n,j,T_{1}}+\xi^{R}P_{n,e}^{R,n,\mathcal{M}\left(e\right),j,T_{2}}\right)\right]\nonumber \\
\label{eq:PnT}
\end{eqnarray}
\vspace{-8mm}

\hrulefill\vspace{-5mm}
\end{figure*}

Furthermore, we simplified the energy dissipation model of~\cite{Auer2010}
in order to formulate the total energy dissipation in macrocell $n$
as~(\ref{eq:PnT}). The effect of the number of TAs, of the energy
dissipation of the RF as well as of the baseband circuits, and the
efficiencies of the power amplifier, feeder cables, cooling system,
mains power supply, and converters has been accounted for in the fixed
energy dissipation terms of $P_{C}^{B}$ and $P_{C}^{R}$, while the
transmit power dependent terms $\xi^{B}$ and $\xi^{R}$ are associated
with the BS $n$ and its RNs, respectively.

Thus, the ESE of macrocell $n$ is given by
\begin{equation}
\eta_{E}^{n}\left(\mathcal{P},\mathcal{S}\right)=\frac{C_{T}^{n}\left(\mathcal{P},\mathcal{S}\right)}{P_{T}^{n}\left(\mathcal{P},\mathcal{S}\right)}.\label{eq:ESEn}
\end{equation}
In the sequel, our aim is to maximize~(\ref{eq:ESEn}) for each macrocell
$n$ by the careful optimization of the variables contained within
$\mathcal{P}$ and $\mathcal{S}$. We define the average ESE of the
multicell system as
\begin{equation}
\eta_{E}\left(\mathcal{P},\mathcal{S}\right)=\frac{1}{3}\sum_{n=1}^{3}\eta_{E}^{n}\left(\mathcal{P},\mathcal{S}\right),\label{eq:ESE}
\end{equation}
so that the average ESE of the system can be optimized by individually
maximizing each macrocell's ESE, as it will be discussed in the following.

\section{Optimization problem formulation and solution algorithm\label{sec:optimization}}

\begin{figure*}
\begin{eqnarray}
\underset{\mathcal{P},\mathcal{S}}{\mbox{maximize}} &  & (\ref{eq:ESE})\label{eq:obj_orig}\\
\mbox{subject to} &  & \sum_{j\in\mathcal{G}^{n}}s^{n,j}\leq1\mbox{, }\forall n,\label{eq:theta_orig}\\
 &  & \sum_{j\in\mathcal{G}^{n}}s^{n,j}\left[\sum_{e_{1}\in\mathcal{E}^{n,j}}P_{n,e_{1}}^{B,n,j,T_{1}}+\sum_{e\in\mathcal{E}^{n,j}}P_{n,e}^{B,n,j,T_{1}}\right]\leq P_{max}^{B}\mbox{, }\forall n,\label{eq:zeta_orig}\\
 &  & \sum_{j\in\mathcal{G}^{n}}s^{n,j}\sum_{e_{2}\in\mathcal{E}^{n,j}}P_{n,e_{2}}^{B,n,j,T_{2}}\leq P_{max}^{B}\mbox{, }\forall n,\label{eq:omega_orig}\\
 &  & \sum_{j\in\mathcal{G}^{n}}s^{n,j}\sum_{\begin{subarray}{c}
e\in\mathcal{E}^{n,j}\\
\mathcal{M}\left(e\right)=m
\end{subarray}}P_{n,e}^{R,n,m,j,T_{2}}\leq P_{max}^{R}\mbox{, }\forall n,m,\label{eq:nu_orig}\\
 &  & s^{n,j}\in\left\{ 0,1\right\} \mbox{, }\forall n,j,\label{eq:sigmatau_orig}\\
 &  & P_{n,e_{1}}^{B,n,j,T_{1}}\mbox{, }P_{n,e}^{B,n,j,T_{1}}\mbox{, }P_{n,e_{2}}^{B,n,j,T_{2}}\mbox{, }P_{n,e}^{R,n,m,j,T_{2}}\geq0\mbox{, }\forall n,j,e_{1},e_{2},e\label{eq:chipsi_orig}
\end{eqnarray}
\hrulefill\vspace{-5mm}
\end{figure*}
In this section, our aim is to optimize the OF~(\ref{eq:ESE}). We
formally describe the optimization problem as~(\ref{eq:obj_orig})--(\ref{eq:chipsi_orig}).
To elaborate,~(\ref{eq:ESE}) is maximized by appropriately optimizing
the decision variables contained within the sets $\mathcal{P}$ and
$\mathcal{S}$. The constraint~(\ref{eq:theta_orig}) ensures that
each macrocell only serves a single SMC group, thus the ICI is completely
avoided. The constraints~(\ref{eq:zeta_orig})--(\ref{eq:nu_orig})
require that none of the transmitters exceeds its maximum transmission
power constraint. Observe that two constraints are needed for each
BS, since each BS transmits in both phases, whereas the RNs only transmit
during the second phase. Furthermore, the constraint~(\ref{eq:sigmatau_orig})
reflects the binary constraint imposed on the $s^{n,j}$ variables,
while the constraints~(\ref{eq:chipsi_orig}) ensures that the power
control variables are non-negative.

\subsection{Concave problem formulation}

Observe that in both the full-IA and partial-IA protocols, the OCI
terms are negligible or zero, if perfect CSI is
available. Therefore, each macrocell's ESE is independent
of the decision variables associated with other macrocells, and the
optimization problem can be decomposed and solved distributively,
where each macrocell optimizes its own ESE. It can be readily proven
that the OF is nonlinear and involves binary variables. Thus, the
optimization problem of~(\ref{eq:obj_orig})--(\ref{eq:chipsi_orig})
is a mixed integer nonlinear programming~(MINLP) problem, which are
typically solved using high-complexity branch-and-bound methods~\cite{Bertsekas1999}.
In order to mitigate the computational burden of finding a solution
to~(\ref{eq:obj_orig})--(\ref{eq:chipsi_orig}), we relax%
\footnote{In~\cite{Yu2006}, such a relaxation results in a time-sharing solution
regarding each subcarrier. In this work, this relaxation may be viewed
as time-sharing of each subcarrier block, as multiple SMC groups can
then occupy a fraction of each subcarrier block in time. Naturally,
the relaxation means that we do not accurately solve the original
problem of~(\ref{eq:obj_orig})--(\ref{eq:chipsi_orig}). In
fact, since we have expanded the space of feasible solutions, solving
the relaxed problem results in an upper bound of the optimal objective
value of the original problem. However, the algorithm devised in this
paper for obtaining the optimal solution to the relaxed problem will
only retain integer values of the relaxed variables.\textbf{ }Therefore,
the algorithm essentially maximizes a lower bound of the relaxed problem.
Having said that, as shown in~\cite{Ng2012c,Ng2012,Cheung2013},
the optimal solution to the original problem is still obtained with
high probability when using the dual decomposition method on the relaxed
problem~(as in this work) as the number of subcarriers tends to infinity.
It was shown that $8$ subcarriers is sufficient for this to be true
in the context of~\cite{Seong2006}, while we have shown that $2$
subcarriers is sufficient in the context of~\cite{Cheung2013}.%
} the binary constraint imposed on the variables $s^{n,j}$ by replacing
the constraint~(\ref{eq:sigmatau_orig}) with
\begin{equation}
0\leq s^{n,j}\leq1\mbox{, }\forall n,j.
\end{equation}
\begin{figure*}
\begin{equation}
t^{n}=\frac{1}{\left(P_{C}^{B}+MP_{C}^{R}\right)+\frac{1}{2}\sum\limits _{j\in\mathcal{G}^{n}}\left[\xi^{B}\sum\limits _{e_{1}\in\mathcal{E}^{n,j}}\widetilde{P}_{n,e_{1}}^{B,n,j,T_{1}}+\sum\limits _{e_{2}\in\mathcal{E}^{n,j}}\widetilde{P}_{n,e_{2}}^{B,n,j,T_{2}}+\sum\limits _{e\in\mathcal{E}^{n,j}}\xi^{B}\widetilde{P}_{n,e}^{B,n,j,T_{1}}+\xi^{R}\widetilde{P}_{n,e}^{R,n,m,j,T_{2}}\right]}\label{eq:tn}
\end{equation}

\hrulefill\vspace{-5mm}
\end{figure*}
Additionally, we introduce the auxiliary variables
\begin{eqnarray}
\widetilde{P}_{n,e_{1}}^{B,n,j,T_{1}} & = & t^{n}s^{n,j}P_{n,e_{1}}^{B,n,j,T_{1}},\\
\widetilde{P}_{n,e}^{B,n,j,T_{1}} & = & t^{n}s^{n,j}P_{n,e}^{B,n,j,T_{1}},\\
\widetilde{P}_{n,e_{2}}^{B,n,j,T_{2}} & = & t^{n}s^{n,j}P_{n,e_{2}}^{B,n,j,T_{2}},\\
\widetilde{P}_{n,e}^{R,n,m,j,T_{2}} & = & t^{n}s^{n,j}P_{n,e}^{R,n,m,j,T_{2}},\\
\widetilde{s}^{n,j} & = & t^{n}s^{n,j}\mbox{, }\forall n,j,e_{1},e_{2},e,
\end{eqnarray}
where $t^{n}$ is given by~(\ref{eq:tn}). Note that we have applied
the Charnes-Cooper variable transformation~\cite{Avriel1988} using
$t^{n}$. Furthermore, the auxiliary SE variables $\widetilde{C}_{n,e_{1}}^{BU,n,j,T_{1}}$,
$\widetilde{C}_{n,e_{2}}^{BU,n,j,T_{2}}$ and $\widetilde{C}_{n,e}^{BRU,n,m,j}$
are introduced, so that we may rewrite the optimization problem of~(\ref{eq:obj_orig})--(\ref{eq:chipsi_orig})
in the hypograph form~\cite{Boyd2004} given by~(\ref{eq:obj_ee})--(\ref{eq:varrho_ee}),
$\forall n$, where $\widetilde{\mathcal{P}}^{n}$, $\widetilde{\mathcal{S}}^{n}$
and $\widetilde{\mathcal{C}}^{n}$ denote the variable sets containing
the auxiliary variables that are associated with macrocell $n$.
\begin{figure*}
\begin{eqnarray}
\underset{\widetilde{\mathcal{P}}^{n},\widetilde{\mathcal{S}}^{n},\widetilde{\mathcal{C}}^{n}}{\mbox{maximize}} &  & \sum_{j\in\mathcal{G}^{n}}\left[\sum_{e_{1}\in\mathcal{E}^{n,j}}\widetilde{C}_{n,k,e_{1}}^{BU,n,j,T_{1}}+\sum_{e_{2}\in\mathcal{E}^{n,j}}\widetilde{C}_{n,k,e_{2}}^{BU,n,j,T_{2}}+\sum_{e\in\mathcal{E}^{n,j}}\widetilde{C}_{n,k,e}^{BRU,n,\mathcal{M}\left(e\right),j}\right]\label{eq:obj_ee}\\
 &  & \frac{\widetilde{s}^{n,j}}{2}\log_{2}\left(1+\frac{w_{n,e_{1}}^{BU,n,j,T_{1}}\widetilde{P}_{n,e_{1}}^{B,n,j,T_{1}}}{\widetilde{s}^{n,j}\Delta\gamma N_{0}LW}\right)\geq\widetilde{C}_{n,e_{1}}^{BU,n,j,T_{1}}\mbox{, }\forall j,e_{1},\label{eq:phi_ee}\\
 &  & \frac{\widetilde{s}^{n,j}}{2}\log_{2}\left(1+\frac{w_{n,e_{2}}^{BU,n,j,T_{2}}\widetilde{P}_{n,e_{2}}^{B,n,j,T_{2}}}{\widetilde{s}^{n,j}\Delta\gamma N_{0}LW}\right)\geq\widetilde{C}_{n,e_{2}}^{BU,n,j,T_{2}}\mbox{, }\forall j,e_{2},\label{eq:varphi_ee}\\
 &  & \frac{\widetilde{s}^{n,j}}{2}\log_{2}\left(1+\frac{w_{n,e}^{BR,n,j,T_{1}}\widetilde{P}_{n,e}^{B,n,j,T_{1}}}{\widetilde{s}^{n,j}\Delta\gamma N_{0}LW}\right)\geq\widetilde{C}_{n,e}^{BRU,n,\mathcal{M}\left(e\right),j}\mbox{, }\forall j,e,\label{eq:kappa_ee}\\
 &  & \frac{\widetilde{s}^{n,j}}{2}\log_{2}\left(1+\frac{w_{n,e}^{RU,n,\mathcal{M}\left(e\right),j,T_{2}}\widetilde{P}_{n,e}^{R,n,\mathcal{M}\left(e\right),j,T_{2}}}{\widetilde{s}^{n,j}\Delta\gamma N_{0}LW}\right)\geq\widetilde{C}_{n,e}^{BRU,n,\mathcal{M}\left(e\right),j}\mbox{, }\forall j,e,\label{eq:varkappa_ee}\\
 &  & \sum_{j\in\mathcal{G}^{n}}\widetilde{s}^{n,j}\leq t^{n},\label{eq:theta_ee}\\
 &  & \sum_{j\in\mathcal{G}^{n}}\left[\sum_{e_{1}\in\mathcal{E}^{n,j}}\widetilde{P}_{n,e_{1}}^{B,n,j,T_{1}}+\sum_{e\in\mathcal{E}^{n,j}}\widetilde{P}_{n,e}^{B,n,j,T_{1}}\right]\leq t^{n}\cdot P_{max}^{B},\label{eq:zeta_ee}\\
 &  & \sum_{j\in\mathcal{G}^{n}}\sum_{e_{2}\in\mathcal{E}^{n,j}}\widetilde{P}_{n,e_{2}}^{B,n,j,T_{2}}\leq t^{n}\cdot P_{max}^{B},\label{eq:omega_ee}\\
 &  & \sum_{j\in\mathcal{G}^{n}}\sum_{\begin{subarray}{c}
e\in\mathcal{E}^{n,j}\\
\mathcal{M}\left(e\right)=m
\end{subarray}}\widetilde{P}_{n,e}^{R,n,m,j,T_{2}}\leq t^{n}\cdot P_{max}^{R}\mbox{, }\forall m,\label{eq:nu_ee}\\
 &  & 0\leq\widetilde{s}^{n,j}\leq t\mbox{, }\forall j,\label{eq:sigmatau_ee}\\
 &  & \widetilde{P}_{n,e_{1}}^{B,n,j,T_{1}},\widetilde{P}_{n,e_{2}}^{B,n,j,T_{2}},\widetilde{P}_{n,e}^{B,n,j,T_{1}}\widetilde{P}_{n,e}^{R,n,m,j,T_{2}}\geq0\mbox{, }\forall j,e_{1},e_{2},e,\label{eq:chipsi_ee}\\
 &  & t^{n}\cdot\left(P_{C}^{B}+M\cdot P_{C}^{R}\right)\nonumber \\
 &  & +\frac{1}{2}\sum_{j\in\mathcal{G}^{n}}\left[\xi^{B}\sum_{e_{1}\in\mathcal{E}^{n,j}}\widetilde{P}_{n,e_{1}}^{B,n,j,T_{1}}+\sum_{e_{2}\in\mathcal{E}^{n,j}}\widetilde{P}_{n,e_{2}}^{B,n,j,T_{2}}+\sum_{e\in\mathcal{E}^{n,j}}\xi^{B}\widetilde{P}_{n,e}^{B,n,j,T_{1}}+\xi^{R}\widetilde{P}_{n,e}^{R,n,m,j,T_{2}}\right]=1\label{eq:varrho_ee}
\end{eqnarray}

\hrulefill\vspace{-5mm}
\end{figure*}
 To elaborate further, the constraints~(\ref{eq:phi_ee}) and~(\ref{eq:varphi_ee})
ensure that the auxiliary SE variables given by $\widetilde{C}_{n,e_{1}}^{BU,n,j,T_{1}}$
and $\widetilde{C}_{n,e_{2}}^{BU,n,j,T_{2}}$ do not exceed the direct
link SEs obtained from~(\ref{eq:c_direct1}) and~(\ref{eq:c_direct2}),
respectively, while the constraints~(\ref{eq:kappa_ee}) and~(\ref{eq:varkappa_ee})
have to be combined to guarantee that~(\ref{eq:c_relayed}) is adhered
to. The constraints~(\ref{eq:theta_ee})--(\ref{eq:chipsi_ee}) are
simply the equivalents of the constraints~(\ref{eq:theta_orig})--(\ref{eq:chipsi_orig}),
when employing the auxiliary variables, while the constraint~(\ref{eq:varrho_ee})
is the result of the Charnes-Cooper variable transformation~\cite{Avriel1988}.
Finally, the OF~(\ref{eq:obj_ee}) defines the ESE of macrocell $n$. 

Let us now aim for proving that~(\ref{eq:obj_ee})--(\ref{eq:varrho_ee})
is a concave maximization problem. It can be readily shown that the
OF~(\ref{eq:obj_ee}) is linear, hence concave. Similarly, the constraints~(\ref{eq:theta_ee})--(\ref{eq:varrho_ee})
are all linear. Therefore, what remains for us to prove is that the
constraints~(\ref{eq:phi_ee})--(\ref{eq:varkappa_ee}) are all convex.
Observe that the constraints~(\ref{eq:phi_ee})--(\ref{eq:varkappa_ee})
are all of the form $\frac{s}{2}\log_{2}\left(1+\frac{aP}{s}\right)\geq C$,
where the decision variables are $s$, $P$ and $C$, while $a$ is
some constant. It is plausible that $\left(1+aP\right)$ is linear.
The function composition of $\frac{1}{2}\log_{2}\left(1+aP\right)$
is concave~\cite{Boyd2004} and the perspective transformation~\cite{Boyd2004},
giving $\frac{s}{2}\log_{2}\left(1+\frac{aP}{s}\right)$, preserves
concavity. Finally, rewriting the previous inequality as $C-\frac{s}{2}\log_{2}\left(1+\frac{aP}{s}\right)\leq0$
clearly shows that it is indeed a convex constraint. Thus, we have
proven that~(\ref{eq:obj_ee})--(\ref{eq:varrho_ee}) is a concave
programming problem, which may be solved using efficient algorithms.
Let us now proceed with the portrayal of the algorithm employed in
this work for solving the above problem.

\subsection{Solution algorithm\label{sub:solution}}

Observe that the optimization problem of~(\ref{eq:obj_ee})--(\ref{eq:varrho_ee})
is akin to a sum-rate maximization problem, which is optimally solved
using the well-known water-filling method~\cite{Goldsmith2005}.
From our previous work~\cite{Cheung2013,Cheung2013a,Cheung2014}
using dual decomposition~\cite{Palomar2006}, we may deduce that
the optimal~(denoted by a superscript asterisk) values for $\widetilde{P}_{n,e_{1}}^{B,n,j,T_{1}}$
and $\widetilde{P}_{n,e_{2}}^{B,n,j,T_{2}}$ are respectively given
by
\begin{equation}
\widetilde{P}_{n,e_{1}}^{B,n,j,T_{1}*}=\widetilde{s}^{n,j}\left[\frac{1}{\left(\xi^{B}\mu^{*}+2\lambda^{n,T_{1}*}\right)\ln2}-\frac{\Delta\gamma N_{0}LW}{w_{n,e_{1}}^{BU,n,j,T_{1}}}\right]^{+}\label{eq:Pbd_T1}
\end{equation}
and
\begin{equation}
\widetilde{P}_{n,e_{2}}^{B,n,j,T_{2}*}=\widetilde{s}^{n,j}\left[\frac{1}{\left(\xi^{B}\mu^{*}+2\lambda^{n,T_{2}*}\right)\ln2}-\frac{\Delta\gamma N_{0}LW}{w_{n,e_{2}}^{BU,n,j,T_{2}}}\right]^{+},\label{eq:Pbd_T2}
\end{equation}
where $\widetilde{s}^{n,j}$ is yet to be determined, while $\left[\cdot\right]^{+}$
is equivalent to $\max\left(0,\cdot\right)$. Furthermore, $\mu^{*}$
is the optimal Lagrangian dual variable~\cite{Boyd2004} associated
with the constraint~(\ref{eq:varrho_ee}), while $\lambda^{n,T_{1}*}$
and $\lambda^{n,T_{2}*}$ are respectively the optimal Lagrangian
dual variables associated with the constraints~(\ref{eq:zeta_ee})
and~(\ref{eq:omega_ee}) for macrocell $n$. The optimal Lagrangian
dual variables are chosen to satisfy the constraints~(\ref{eq:zeta_ee})--(\ref{eq:nu_ee})
with equality, and are found using the subgradient algorithm~\cite{Palomar2006}.

It may be shown that the power control variables of the relaying links
may be formulated as
\begin{equation}
\widetilde{P}_{n,e}^{B,n,j,T_{1}}=\widetilde{s}^{n,j}\left[\frac{1}{\left(\xi^{B}\mu^{*}+2\lambda^{n,T_{1}*}\right)\ln2}-\frac{\Delta\gamma N_{0}LW}{w_{n,e}^{BR,n,j,T_{1}}}\right]^{+}
\end{equation}
and
\begin{eqnarray}
\widetilde{P}_{n,e}^{R,n,\mathcal{M}\left(e\right),j,T_{2}} & =\widetilde{s}^{n,j} & \left[\frac{1}{\left(\xi^{R}\mu^{*}+2\nu^{n,\mathcal{M}\left(e\right),T_{2}*}\right)\ln2}\right.\nonumber \\
 &  & \left.-\frac{\Delta\gamma N_{0}LW}{w_{n,e}^{RU,n,\mathcal{M}\left(e\right),j,T_{2}}}\right]^{+},
\end{eqnarray}
where $\nu^{n,\mathcal{M}\left(e\right),T_{2}*}$ is the optimal Lagrangian
dual variable associated with the constraint~(\ref{eq:nu_ee}) for
RN $\mathcal{M}\left(e\right)$ belonging to macrocell $n$. Since
the attainable SE of a relaying link is limited by the weaker of the
BS-RN and RN-UE links, there is no need to transmit at a higher power
than necessary, if the other link is unable to support the higher
SE. Thus, the optimal power control variables for the relaying link
are given by
\begin{eqnarray}
\widetilde{P}_{n,e}^{B,n,j,T_{1}*} & = & \min\Bigg(\frac{w_{n,e}^{RU,n,\mathcal{M}\left(e\right),j,T_{2}}}{w_{n,e}^{BR,n,j,T_{1}}}\cdot\widetilde{P}_{n,e}^{R,n,\mathcal{M}\left(e\right),j,T_{2}},\nonumber \\
 &  & \widetilde{P}_{n,e}^{B,n,j,T_{1}}\Bigg)\label{eq:Pbr_T1}
\end{eqnarray}
and
\begin{eqnarray}
\widetilde{P}_{n,e}^{R,n,\mathcal{M}\left(e\right),j,T_{2}*} & = & \min\Bigg(\frac{w_{n,e}^{BR,n,j,T_{1}}}{w_{n,e}^{RU,n,\mathcal{M}\left(e\right),j,T_{2}}}\cdot\widetilde{P}_{n,e}^{B,n,j,T_{1}},\nonumber \\
 &  & \widetilde{P}_{n,e}^{R,n,\mathcal{M}\left(e\right),j,T_{2}}\Bigg).\label{eq:Pbr_T2}
\end{eqnarray}
\begin{figure*}
\begin{equation}
\widetilde{C}_{n,e}^{BRU,n,\mathcal{M}\left(e\right),j}=\frac{\widetilde{s}^{n,j}}{2}\log_{2}\left(1+\frac{w_{n,e}^{BR,n,j,T_{1}}\widetilde{P}_{n,e}^{B,n,j,T_{1}}}{\Delta\gamma N_{0}LW}\right)=\frac{\widetilde{s}^{n,j}}{2}\log_{2}\left(1+\frac{w_{n,e}^{RU,n,\mathcal{M}\left(e\right),j,T_{2}}\widetilde{P}_{n,e}^{R,n,\mathcal{M}\left(e\right),j,T_{2}}}{\Delta\gamma N_{0}LW}\right)\label{eq:Cbru}
\end{equation}

\hrulefill
\end{figure*}
Thus, the maximum values of $\widetilde{C}_{n,e_{1}}^{BU,n,j,T_{1}}$,
$\widetilde{C}_{n,e_{2}}^{BU,n,j,T_{2}}$ and $\widetilde{C}_{n,e}^{BRU,n,\mathcal{M}\left(e\right),j}$
are given by
\begin{equation}
\widetilde{C}_{n,e}^{BU,n,j,T_{1}}=\frac{\widetilde{s}^{n,j}}{2}\log_{2}\left(1+\frac{w_{n,e}^{BU,n,j,T_{1}}\widetilde{P}_{n,e}^{B,n,j,T_{1}}}{\Delta\gamma N_{0}LW}\right),\label{eq:Cb_T1}
\end{equation}
\begin{equation}
\widetilde{C}_{n,e}^{BU,n,j,T_{2}}=\frac{\widetilde{s}^{n,j}}{2}\log_{2}\left(1+\frac{w_{n,e}^{BU,n,j,T_{2}}\widetilde{P}_{n,e}^{B,n,j,T_{2}}}{\Delta\gamma N_{0}LW}\right)\label{eq:Cb_T2}
\end{equation}
and~(\ref{eq:Cbru}), where the value of $\widetilde{s}^{n,j}$ is
not yet known. However, regardless of the exact value of $\widetilde{s}^{n,j}$,
macrocell $n$ may choose the specific SMC group $j$ that obtains
the highest value of
\begin{eqnarray}
 &  & \sum_{e_{1}\in\mathcal{E}^{n,j}}\widetilde{C}_{n,e_{1}}^{BU,n,j,T_{1}}\nonumber \\
 &  & +\sum_{e_{2}\in\mathcal{E}^{n,j}}\widetilde{C}_{n,e_{2}}^{BU,n,j,T_{2}}+\sum_{e\in\mathcal{E}^{n,j}}\widetilde{C}_{n,e}^{BRU,n,\mathcal{M}\left(e\right),j}\label{eq:C_max}
\end{eqnarray}
in order to maximize the OF~(\ref{eq:obj_ee}) by setting $\widetilde{s}^{n,j}=t^{n}$,
where the value of $t^{n}$ is not yet known. As a result, the SMC
groups $j'\neq j$ are not chosen and we may set $\widetilde{s}^{n,j'}=\widetilde{P}_{n,e_{1}}^{B,n,j',T_{1}}=\widetilde{P}_{n,e_{2}}^{B,n,j',T_{2}}=\widetilde{P}_{n,e}^{B,n,j',T_{1}}=\widetilde{P}_{n,e}^{R,n,\mathcal{M}\left(e\right),j',T_{2}}=\widetilde{C}_{n,e_{1}}^{BU,n,j',T_{1}}=\widetilde{C}_{n,e_{2}}^{BU,n,j',T_{2}}=\widetilde{C}_{n,e}^{BRU,n,\mathcal{M}\left(e\right),j'}=0$,
$\forall e_{1},e_{2},e,j'\neq j$.

\begin{figure*}
\begin{equation}
t^{n}=\left(P_{C}^{B}+M\cdot P_{C}^{R}+\sum_{j\in\mathcal{G}^{n}}\frac{1}{2\widetilde{s}^{n,j}}\left[\xi^{B}\sum_{e_{1}\in\mathcal{E}^{n,j}}\widetilde{P}_{n,e_{1}}^{B,n,j,T_{1}}+\sum_{e_{2}\in\mathcal{E}^{n,j}}\widetilde{P}_{n,e_{2}}^{B,n,j,T_{2}}+\sum_{e\in\mathcal{E}^{n,j}}\xi^{B}\widetilde{P}_{n,e}^{B,n,j,T_{1}}+\xi^{R}\widetilde{P}_{n,e}^{R,n,m,j,T_{2}}\right]\right)^{-1}\label{eq:optT}
\end{equation}
\hrulefill\vspace{-5mm}
\end{figure*}
The optimal value of $t^{n}$ is then given by~(\ref{eq:optT}).
Observe that this is possible, since~(\ref{eq:optT}) is only dependent
on the dual variables. Furthermore, determining the value of $t^{n}$
gives the values of $\widetilde{s}^{n,j}$, $\widetilde{P}_{n,e_{1}}^{B,n,j,T_{1}}$,
$\widetilde{P}_{n,e_{2}}^{B,n,j,T_{2}}$, $\widetilde{P}_{n,e}^{B,n,j,T_{1}}$
and $\widetilde{P}_{n,e}^{R,n,m,j,T_{2}}$ for the selected SMC group.

By following the above derivations, the constraints~(\ref{eq:phi_ee})--(\ref{eq:theta_ee})
and~(\ref{eq:sigmatau_ee})--(\ref{eq:varrho_ee}) are implicitly
satisfied and there is no need to introduce dual variables for them.
This ESEM solution algorithm may be implemented distributively, and
iterates between obtaining the optimal primal variables and applying
the subgradient method~\cite{Palomar2006} for updating the dual
variables, until the change in the dual variable values becomes less
than $\epsilon$ or the maximum number of iterations, $I_{max}$,
has been reached. The ESEM algorithm is summarized in Table~\ref{algor:optimization},
where $\lambda^{n,T_{1}}\left(i\right)$, $\lambda^{n,T_{2}}\left(i\right)$,
$\nu^{n,m,T_{2}}\left(i\right)$ and $\mu\left(i\right)$ indicate
the value of their respective dual variables at the $i$th iteration.
\begin{table}
\setlength{\arrayrulewidth}{1pt}

\centering{}\caption{The ESEM algorithm based on dual decomposition and the subgradient
method.}
\label{algor:optimization}%
\begin{tabular}{rl}
\hline 
\multicolumn{2}{l}{\textbf{Algorithm 1} ESEM algorithm}\tabularnewline
\hline 
1: & $i\leftarrow0$\tabularnewline
2: & \textbf{do while} $|\lambda^{n,T_{1}}\left(i\right)-\lambda^{n,T_{1}}\left(i-1\right)|>\epsilon$
\textbf{or} \tabularnewline
 & $|\lambda^{n,T_{1}}\left(i\right)-\lambda^{n,T_{1}}\left(i-1\right)|>\epsilon$
\textbf{or} \tabularnewline
 & $|\nu^{n,m,T_{2}}\left(i\right)-\nu^{n,m,T_{2}}\left(i-1\right)|>\epsilon$
\textbf{or} \tabularnewline
 & $|\mu\left(i\right)-\mu\left(i-1\right)|>\epsilon$\tabularnewline
3: & \quad{}$i\leftarrow i+1$\tabularnewline
4: & \quad{}\textbf{if} $i>I_{max}$\tabularnewline
5: & \quad{}\quad{}\textbf{break}\tabularnewline
6: & \quad{}\textbf{end if}\tabularnewline
7: & \quad{}\textbf{for} $n$ \textbf{from} $1$ \textbf{to} $3$\tabularnewline
8: & \quad{}\quad{}\textbf{for each} $j$ \textbf{in} $\mathcal{G}^{n}$\tabularnewline
9: & \quad{}\quad{}\quad{}Obtain the optimal power allocation using~(\ref{eq:Pbd_T1})--(\ref{eq:Pbr_T2})\tabularnewline
10: & \quad{}\quad{}\quad{}Compute their achievable SE using~(\ref{eq:Cb_T1})--(\ref{eq:Cbru})\tabularnewline
11: & \quad{}\quad{}\textbf{end for}\tabularnewline
12: & \quad{}\quad{}Find the optimal SMC, which obtains the maximum~(\ref{eq:C_max})\tabularnewline
13: & \quad{}\quad{}Compute the optimal $t$ using~(\ref{eq:optT})\tabularnewline
14: & \quad{}\textbf{end for}\tabularnewline
15: & \quad{}Update the dual variables $\lambda^{n,T_{1}}\left(i\right)$,
$\lambda^{n,T_{2}}\left(i\right)$, $\nu^{n,m,T_{2}}\left(i\right)$\tabularnewline
 & \quad{}and $\mu\left(i\right)$ using the subgradient method~\cite{Palomar2006}\tabularnewline
16: & \textbf{end do}\tabularnewline
17: & \textbf{return}\tabularnewline
\hline 
\end{tabular}
\end{table}

\section{Numerical results and discussions\label{sec:results}}

\begin{table}[t]
\begin{centering}
\caption{Simulation parameters used to obtain all results in this section unless
otherwise specified.}
\label{tab:param}\setlength{\extrarowheight}{3pt}
\par\end{centering}

\centering{}%
\begin{tabular}{|l|r|}
\hline 
Simulation parameter & Value\tabularnewline
\hline 
\hline 
Subcarrier block bandwidth, $W$ {[}Hertz{]} & $180$k\tabularnewline
\hline 
Number of RNs per macrocell, $M$ & $\{0,1,2,3\}$\tabularnewline
\hline 
Number of subcarriers blocks, $N$ & $12$\tabularnewline
\hline 
Number of UEs, $K$ & $6$\tabularnewline
\hline 
Antenna configuration, $\left(N_{B},N_{R},N_{U}\right)$ & $\left(4,4,4\right)$\tabularnewline
\hline 
Semi-orthogonality parameter, $\alpha$ & $0.1$\tabularnewline
\hline 
Inter-site distance~(ISD), {[}km{]} & $\{1.5,2.5,3.5,4.5\}$\tabularnewline
\hline 
Minimum number of receive dimensions & \tabularnewline
at the RNs and UEs, $S^{R}$ and $S^{U}$ & $1$ and $2$\tabularnewline
\hline 
Ratio of BS-to-RN distance to the cell & \tabularnewline
radius, $D_{r}$ & $0.7$\tabularnewline
\hline 
SNR gap of wireless transceivers, $\Delta\gamma$ {[}dB{]} & 0\tabularnewline
\hline 
\multicolumn{1}{|l|}{Maximum total transmission power of the} & $\{0,12,24,$\tabularnewline
BS and RNs, $P_{max}^{B}$ and $P_{max}^{R}$ {[}dBm{]} & $36,48,60\}$\tabularnewline
\hline 
Fixed power rating of the BS, & 32.306$N_{B}$\tabularnewline
$P_{C}^{B}$ {[}Watts{]}~\cite{Arnold2010,Auer2010} & \tabularnewline
\hline 
Fixed power rating of RNs, & 21.874$N_{R}$\tabularnewline
$P_{C}^{R}$ {[}Watts{]}~\cite{Arnold2010,Auer2010} & \tabularnewline
\hline 
Reciprocal of the BS power amplifier's & 3.24$N_{B}$\tabularnewline
drain efficiency, $\xi^{B}$~\cite{Arnold2010,Auer2010} & \tabularnewline
\hline 
Reciprocal of the RNs' power amplifier's & 4.04$N_{R}$\tabularnewline
drain efficiency, $\xi^{R}$~\cite{Arnold2010,Auer2010} & \tabularnewline
\hline 
Noise power spectral density, $N_{0}$ {[}dBm/Hz{]} & \textminus{}174\tabularnewline
\hline 
Convergence threshold, $\epsilon$ & $10^{-8}$\tabularnewline
\hline 
Number of channel samples & $10^{4}$\tabularnewline
\hline 
\end{tabular}\vspace{-5mm}
\end{table}
This section presents the numerical results%
\footnote{In all cases, the step sizes and the initial values of the dual variables
described in Section~\ref{sub:solution} are empirically optimized
so that the algorithm converges in as few iterations as possible,
although the exact analytical method for achieving this still remains
an open issue. In our experience, the algorithm converges within just
$10$ iterations when carefully chosen step sizes are employed, regardless
of the size of the problem.%
} obtained, when the solution algorithm presented in Section~\ref{sub:solution}
is employed for the ESEM problem of~(\ref{eq:obj_ee})--(\ref{eq:varrho_ee}),
where the simulation parameters are given in Table~\ref{tab:param}.
Furthermore, we employed the path-loss model of~\cite{3GPP_PL} and
assumed that all BS-UE and RN-UE links are NLOS links, since they
are typically blocked by buildings and other large obstructing objects,
while all BS-RN links may realistically be assumed to be line-of-sight
links, since the RNs may be strategically positioned on tall buildings
to create strong wireless backhaul links. Furthermore, independently
and randomly generated set of UE locations as well as fading channel
realizations were used. Again, for benchmarking we employ a baseline
algorithm, which relies on random SMC selections and equal power allocation
across the selected SMCs. This algorithm is termed as the EPA algorithm.

The attainable performance of both the full-IA and partial-IA protocols
is explored and these results are obtained by employing the optimized
power control variables and group selection variables in the actual
system model. Therefore, the results reflect the actual ESE achieved
rather than the optimized OF value of~(\ref{eq:obj_ee}), which is
optimistic, since it does not account for any potential OCI remaining
after employing the partial-IA protocol.

\subsection{The variation of ASE and ESE for different values of $P_{max}^{B}$
and $P_{max}^{R}$}

\begin{figure}
\begin{centering}
\subfloat[Surface plots of the achievable ASE when using the ESEM and EPA algorithms.]{\begin{centering}
\includegraphics[scale=0.73]{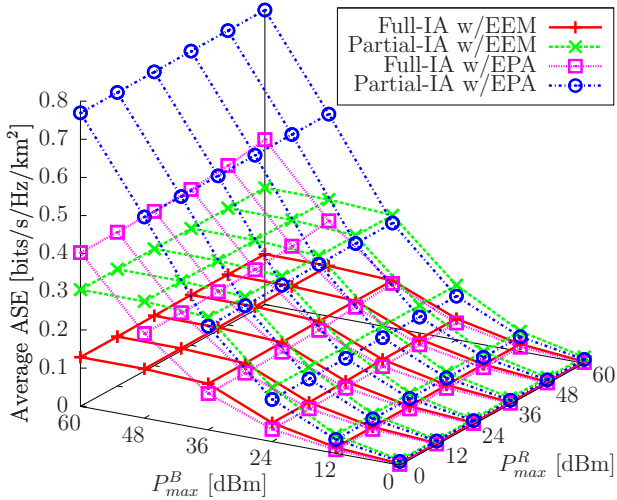}
\par\end{centering}

}
\par\end{centering}

\begin{centering}
\subfloat[Surface plots of the achievable ESE when using the ESEM and EPA algorithms.]{\begin{centering}
\includegraphics[scale=0.73]{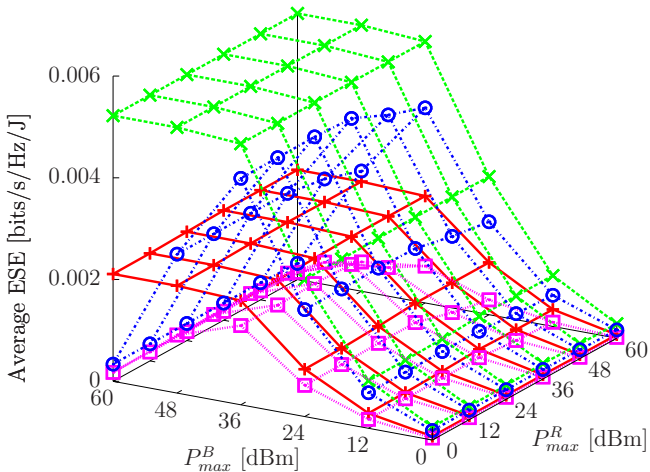}
\par\end{centering}

}
\par\end{centering}

\caption{The average achievable ASE and ESE when using the ESEM and EPA algorithms
with either full-IA or partial-IA, for varying $P_{max}^{B}$ and
$P_{max}^{R}$, and using the parameters in Table~\ref{tab:param}
with $M=2$ and an ISD of $1.5$km.}
\label{fig:powers}\vspace{-5mm}
\end{figure}
The effects of varying both $P_{max}^{B}$ and $P_{max}^{R}$
are demonstrated in Fig.~\ref{fig:powers}. Observe that the partial-IA
protocol outperforms the full-IA protocol for all the power constraints
considered. This is due to the requirements of~(\ref{eq:full_ia_b1}),~(\ref{eq:partial_ia_b1}),~(\ref{eq:full_ia_b2})
and~(\ref{eq:partial_ia_b2}), which restrict the number of data
streams that the BSs can transmit simultaneously in each phase. The
full-IA protocol imposes more restrictive constraints than the partial-IA
protocol, since the partial-IA protocol only requires that the Rx-BFMs
has to eliminate the ICI, rather than both the ICI and OCI that the
full-IA protocol has to null. Observe furthermore that the EPA algorithms
achieve higher ASE values than their ESEM algorithmic counterparts
at high $P_{max}^{B}$ values. However, this is achieved at a higher
cost to the ESE obtained from using the EPA algorithms, when compared
to their ESEM counterparts. In fact, in the low to medium $P_{max}^{B}$
regime, both the SEM and ESEM correspond to the same solution, as
demonstrated in our previous works of~\cite{Cheung2013,Cheung2013a,Cheung2014}.
This results in a higher ASE for the ESEM algorithm than for the heuristic
EPA algorithm. As the value of $P_{max}^{B}$ increases, the EPA continues
to allocate more power, which increases the ASE obtained, without
any cognizance to the ESE performance.

However, the ASE and ESE obtained does not increase significantly
upon increasing $P_{max}^{R}$. This can be attributed to the low
multiplexing gain specified in these experiments, given that $S^{R}=1$.
The results of the next subsection explore the effects of varying
the requirements imposed on $S^{U}$ and $S^{R}$.

\subsection{The variation of ASE and ESE for different values of $S^{U}$ and
$S^{R}$}

\begin{figure}
\begin{centering}
\subfloat[Surface plots of the achievable ASE when using the ESEM and EPA algorithms.]{\begin{centering}
\includegraphics[scale=0.73]{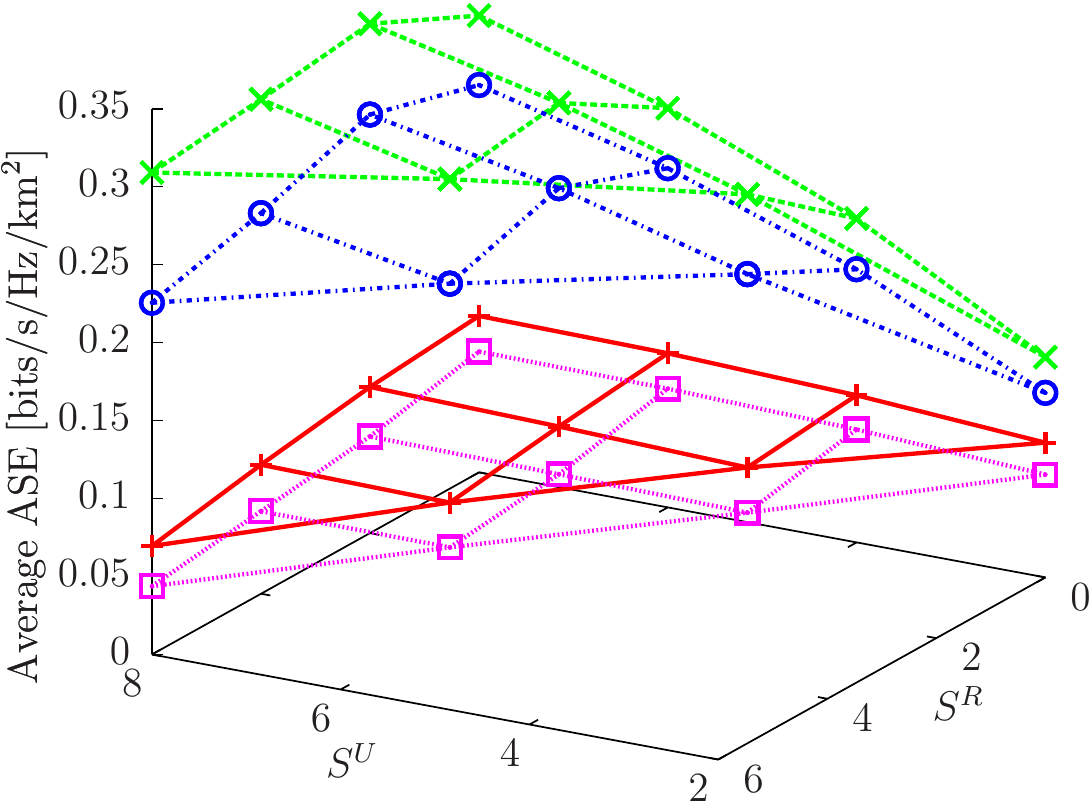}
\par\end{centering}

}
\par\end{centering}

\begin{centering}
\subfloat[Surface plots of the achievable ESE when using the ESEM and EPA algorithms.]{\begin{centering}
\includegraphics[scale=0.73]{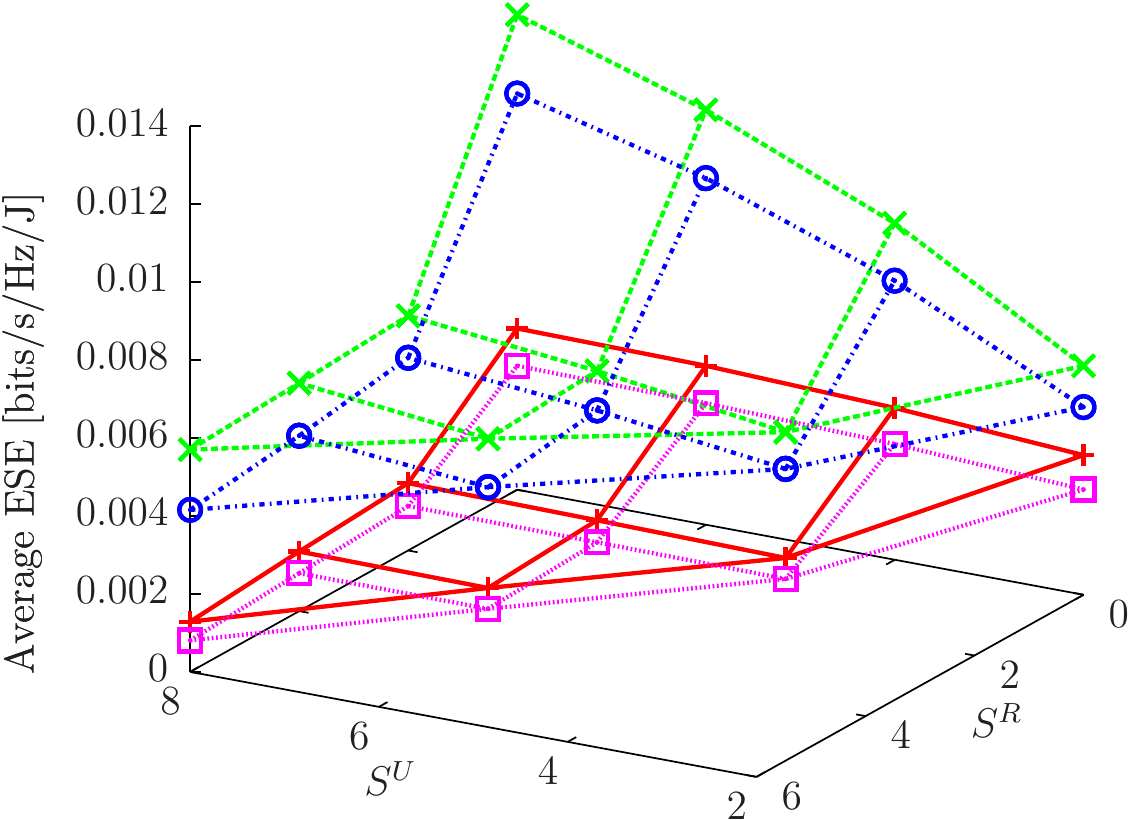}
\par\end{centering}

}
\par\end{centering}

\caption{The average achievable ASE and ESE when using the ESEM and EPA algorithms
with either full-IA or partial-IA, for varying $S^{U}$ and $S^{R}$,
and using the parameters in Table~\ref{tab:param} with $M=2$, $P_{max}^{B}=30$dBm,
$P_{max}^{R}=20$dBm and an ISD of $1.5$km. The legend is as presented
in Fig.~\ref{fig:powers}.}
\label{fig:complexity}\vspace{-5mm}
\end{figure}
Fig.~\ref{fig:complexity} shows the results obtained
upon varying $S^{U}$ and $S^{R}$. Once again, the partial-IA protocol
outperforms the full-IA protocol in terms of both its ASE and ESE
performances. Additionally, we observe that the EPA algorithm performs
worse than the ESEM algorithm for all cases. Increasing $S^{U}$ has
a marginal effect on the ASE and ESE obtained for both protocols.
However, increasing $S^{R}$ does lead to an increase in SE, when
employing the partial-IA protocol, albeit at a cost to ESE resulting
from the fixed power dissipation costs of the RNs. Observe that increasing
$S^{R}$ reduces the ASE attained when using the full-IA protocol.
This may be explained by the detrimental effects of the constraints
imposed on the multiplexing gain of the BSs' transmissions when employing
the full-IA protocol, because increasing $S^{R}$ imposes a substantial
reduction on both~(\ref{eq:full_ia_b1}) and~(\ref{eq:full_ia_b2}),
when multiple RNs are operated in each macrocell. This reduction in
ASE is not so dominant for the partial-IA protocol, since the increase
in the multiplexing gain of the RNs' transmissions outweighs the detrimental
effects of imposing a multiplexing gain restriction at the BSs due
to~(\ref{eq:partial_ia_b2}). Additionally, the potential multiplexing
gain attained at the BSs in the first transmission phase, given by~(\ref{eq:partial_ia_b1}),
is not affected by the increase of $S^{R}$.

\subsection{The variation of ASE and ESE for different values of $M$ and inter-site
distance}

\begin{figure}
\begin{centering}
\subfloat[Surface plots of the achievable ASE when using the ESEM and EPA algorithms.]{\begin{centering}
\includegraphics[scale=0.73]{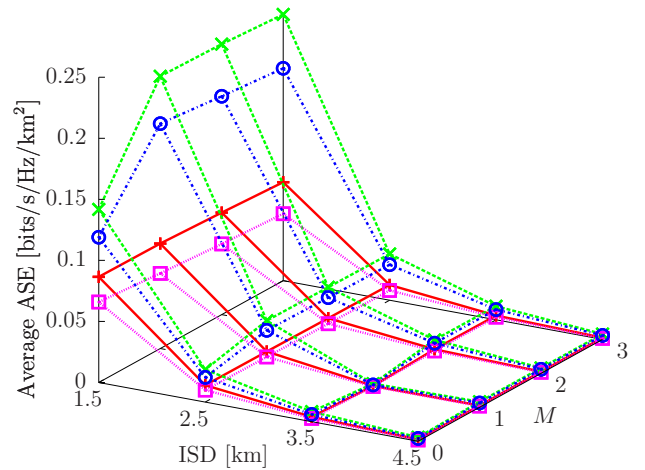}
\par\end{centering}

}
\par\end{centering}

\begin{centering}
\subfloat[Surface plots of the achievable ESE when using the ESEM and EPA algorithms.]{\begin{centering}
\includegraphics[scale=0.73]{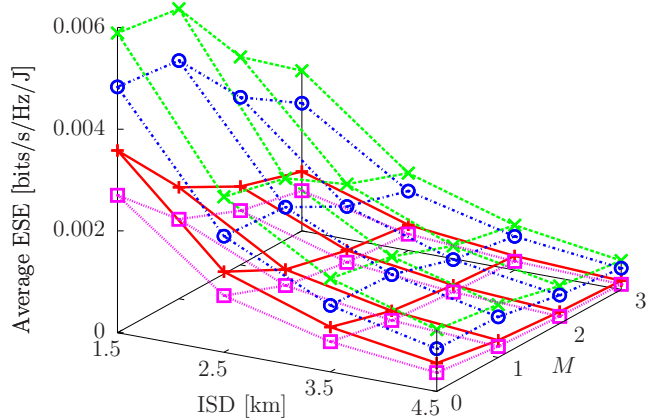}
\par\end{centering}

}
\par\end{centering}

\caption{The average achievable ASE and ESE when using the ESEM and EPA algorithms
with either full-IA or partial-IA, for varying $M$ and ISD, and using
the parameters in Table~\ref{tab:param} with $P_{max}^{B}=30$dBm,
$P_{max}^{R}=20$dBm and an ISD of $1.5$km. The legend is as presented
in Fig.~\ref{fig:powers}.}
\label{fig:relays}\vspace{-5mm}
\end{figure}
As shown in Fig.~\ref{fig:relays}, both the achievable
ASE and ESE decreases as the ISD is increased, indicating that the
effect of a higher path-loss on the channel gains has a more grave
detrimental effect on both the ASE and ESE than the beneficial effects
of the reduced interference levels. Once again, the EPA algorithm
performs worse than their ESEM algorithmic counterparts. Additionally,
the ASE attained, when using the full-IA protocol is slightly reduced
upon increasing $M$ due to both~(\ref{eq:full_ia_b1}) and~(\ref{eq:full_ia_b2}),
while the ESE achieved is reduced, as the power dissipation of the
system is increased upon increasing $M$. Furthermore, the ASE obtained
when using the partial-IA protocol peaks for $M=1$, but decreases
slightly, upon increasing $M$ further, since then the multiplexing
gain of the experienced during the second phase is reduced as indicated
by~(\ref{eq:partial_ia_b2}). By contrast, the ESEM of the partial-IA
protocol only decreases upon increasing $M$%
\footnote{In fact, when $M=0$ or $S^{R}=0$ we arrive at
a special case of the partial-IA protocol, which is similar to the
conventional single cell multi-user ZFBF in the absence of RNs. However,
the proposed partial-IA protocol represents a sophisticated extension
of classic ZFBF to the broad class of multi-relay aided multi-cell
networks, which have been combined with intelligent user selection.}.

\section{Conclusions\label{sec:conc}}

In this paper, a multi-user, multi-relay, multi-cell MIMO system model
is studied. In order to avoid the excessive interference inflicted
by the multiple transmission sources, a pair of distributed IA protocols
were designed. The first, termed as full-IA, completely avoids any
interference by finding RxBFMs, which entirely eliminate the interference
imposed at the receivers. However, this comes at a cost to the spatial
multiplexing gain of the BSs, which limits the number of DL transmission
streams. The second transmission protocol, namely partial-IA, aims
for striking a balance between the spatial multiplexing gain and interference
contamination by finding RxBFMs, which only null the interference
emerging from sources within the same macrocell. Employing the RxBFMs
created by either of these transmission protocols results in a list
of SMCs, which correspond to data streams that may be conveyed by
the BS. We formally defined the problem of maximizing the ESE by optimally
choosing the SMCs as well as by appropriately choosing their power
control variables. The resultant non-convex optimization problem was
converted into a convex optimization problem with the aid of carefully
chosen variable relaxations and transformations, which was then solved
using the classic dual decomposition and subgradient methods~\cite{Palomar2006},
that may be implemented distributively at each BS. We characterized
the attainable ASE and ESE performances of both protocols for a range
of system parameters, while comparing the performance of our ESEM
algorithm to that of a baseline EPA algorithm. To summarize, the ESEM
algorithm outperforms the EPA algorithm in terms
of ESE, while surprisingly the partial-IA protocol
outperforms the full-IA protocol in all cases. For the cell sizes
considered, the path-loss mitigates the majority of the OCI, and thus
the full-IA protocol actually over-compensates, when reducing the
number of available transmit dimensions at the transmitters to facilitate
IA.\vspace{-5mm}

\bibliographystyle{IEEEtran}
\bibliography{references}

\begin{thebibliography}{10}
\providecommand{\url}[1]{#1}
\csname url@samestyle\endcsname
\providecommand{\newblock}{\relax}
\providecommand{\bibinfo}[2]{#2}
\providecommand{\BIBentrySTDinterwordspacing}{\spaceskip=0pt\relax}
\providecommand{\BIBentryALTinterwordstretchfactor}{4}
\providecommand{\BIBentryALTinterwordspacing}{\spaceskip=\fontdimen2\font plus
\BIBentryALTinterwordstretchfactor\fontdimen3\font minus
  \fontdimen4\font\relax}
\providecommand{\BIBforeignlanguage}[2]{{%
\expandafter\ifx\csname l@#1\endcsname\relax
\typeout{** WARNING: IEEEtran.bst: No hyphenation pattern has been}%
\typeout{** loaded for the language `#1'. Using the pattern for}%
\typeout{** the default language instead.}%
\else
\language=\csname l@#1\endcsname
\fi
#2}}
\providecommand{\BIBdecl}{\relax}
\BIBdecl

\bibitem{Lopez-Perez2009}
D.~Lopez-Perez, A.~Valcarce, G.~de~la Roche, and J.~Zhang, ``{OFDMA}
  femtocells: A roadmap on interference avoidance,'' \emph{IEEE Communications
  Magazine}, vol.~47, no.~9, pp. 41--48, Sept. 2009.

\bibitem{Bhat2012}
P.~Bhat, S.~Nagata, L.~Campoy, I.~Berberana, T.~Derham, G.~Liu, X.~Shen,
  P.~Zong, and J.~Yang, ``{LTE}-advanced: an operator perspective,'' \emph{IEEE
  Communications Magazine}, vol.~50, no.~2, pp. 104--114, Feb. 2012.

\bibitem{El-Ayach2013}
O.~El~Ayach, S.~Peters, and R.~Heath~Jr., ``The practical challenges of
  interference alignment,'' \emph{IEEE Wireless Communications Magazine},
  vol.~20, no.~1, pp. 35--42, Feb. 2013.

\bibitem{Han2011}
C.~Han, T.~Harrold, S.~Armour, I.~Krikidis, S.~Videv, P.~Grant, H.~Haas,
  J.~Thompson, I.~Ku, C.-X. Wang, T.~A. Le, M.~Nakhai, J.~Zhang, and L.~Hanzo,
  ``Green radio: radio techniques to enable energy-efficient wireless
  networks,'' \emph{IEEE Communications Magazine}, vol.~49, no.~6, pp. 46--54,
  Jun. 2011.

\bibitem{Laneman2004}
J.~Laneman, D.~Tse, and G.~Wornell, ``Cooperative diversity in wireless
  networks: Efficient protocols and outage behavior,'' \emph{IEEE Transactions
  on Information Theory}, vol.~50, no.~12, pp. 3062--3080, Dec. 2004.

\bibitem{Maddah-Ali2006a}
M.~Maddah-Ali, A.~Motahari, and A.~Khandani, ``Communication over {X} channel:
  Signalling and multiplexing gain,'' \emph{Technical Report. UW-ECE-2006-12,
  University of Waterloo}, Jul. 2006.

\bibitem{Maddah-Ali2006b}
------, ``Communication over {X} channel: Signaling and performance analysis,''
  \emph{Technical Report. UW-ECE-2006-27, University of Waterloo}, Dec. 2006.

\bibitem{Maddah-Ali2006}
------, ``Signaling over {MIMO} multi-base systems: Combination of multi-access
  and broadcast schemes,'' in \emph{Proceedings of the IEEE International
  Symposium on Information Theory}, Seattle, WA, USA, Jul. 2006, pp.
  2104--2108.

\bibitem{Maddah-Ali2008}
------, ``Communication over {MIMO} {X} channels: Interference alignment,
  decomposition, and performance analysis,'' \emph{IEEE Transactions on
  Information Theory}, vol.~54, no.~8, pp. 3457--3470, Aug. 2008.

\bibitem{Cadambe2008}
V.~Cadambe and S.~Jafar, ``Interference alignment and degrees of freedom of the
  $k$-user interference channel,'' \emph{IEEE Transactions on Information
  Theory}, vol.~54, no.~8, pp. 3425--3441, Aug. 2008.

\bibitem{Suh2008}
C.~Suh and D.~Tse, ``Interference alignment for cellular networks,'' in
  \emph{Allerton Conference on Communication, Control, and Computing},
  Urbana-Champaign, IL, USA, Sept. 2008, pp. 1037--1044.

\bibitem{Gao2014}
H.~Gao, T.~Lv, D.~Fang, S.~Yang, and C.~Yuen, ``Limited feedback-based
  interference alignment for interfering multi-access channels,'' \emph{IEEE
  Communications Letters}, vol.~18, no.~4, pp. 540--543, Apr. 2014.

\bibitem{Suh2011}
C.~Suh, M.~Ho, and D.~Tse, ``Downlink interference alignment,'' \emph{IEEE
  Transactions on Communications}, vol.~59, no.~9, pp. 2616--2626, Sept. 2011.

\bibitem{Kim2010}
D.~Kim and M.~Torlak, ``Optimization of interference alignment beamforming
  vectors,'' \emph{IEEE Journal on Selected Areas in Communications}, vol.~28,
  no.~9, pp. 1425--1434, Dec. 2010.

\bibitem{Da2011}
B.~Da and R.~Zhang, ``Exploiting interference alignment in multi-cell
  cooperative {OFDMA} resource allocation,'' in \emph{Proceedings of the IEEE
  Global Telecommunications Conference}, Houston, TX, USA, Dec. 2011.

\bibitem{Gomadam2011}
K.~Gomadam, V.~Cadambe, and S.~Jafar, ``A distributed numerical approach to
  interference alignment and applications to wireless interference networks,''
  \emph{IEEE Transactions on Information Theory}, vol.~57, no.~6, pp.
  3309--3322, Jun. 2011.

\bibitem{Rezaee2012}
M.~Rezaee and S.~Nader-Esfahani, ``Interference alignment for downlink
  transmission of multiple interfering cells,'' \emph{IEEE Wireless
  Communications Letters}, vol.~1, no.~5, pp. 460--463, Oct. 2012.

\bibitem{Tang2013}
J.~Tang and S.~Lambotharan, ``Interference alignment techniques for {MIMO}
  multi-cell interfering broadcast channels,'' \emph{IEEE Transactions on
  Communications}, vol.~61, no.~1, pp. 164--175, Feb. 2013.

\bibitem{Yang2013}
\BIBentryALTinterwordspacing
H.~J. Yang, W.-Y. Shin, B.~C. Jung, C.~Suh, and A.~Paulraj, ``Opportunistic
  downlink interference alignment.'' [Online]. Available:
  \url{http://arxiv.org/abs/1312.7198}
\BIBentrySTDinterwordspacing

\bibitem{Yoo2006}
T.~Yoo and A.~Goldsmith, ``On the optimality of multiantenna broadcast
  scheduling using zero-forcing beamforming,'' \emph{IEEE Journal on Selected
  Areas in Communications}, vol.~24, no.~3, pp. 528--541, Mar. 2006.

\bibitem{Goldsmith2005}
A.~Goldsmith, \emph{Wireless Communications}.\hskip 1em plus 0.5em minus
  0.4em\relax Cambridge University Press, New York, NY, USA, 2005.

\bibitem{Xiong2012}
C.~Xiong, G.~Li, S.~Zhang, Y.~Chen, and S.~Xu, ``Energy-efficient resource
  allocation in {OFDMA} networks,'' \emph{IEEE Transactions on Communications},
  vol.~60, no.~12, pp. 3767--3778, Dec. 2012.

\bibitem{Ng2012a}
D.~Ng, E.~Lo, and R.~Schober, ``Energy-efficient resource allocation in
  multi-cell {OFDMA} systems with limited backhaul capacity,'' \emph{IEEE
  Transactions on Wireless Communications}, vol.~11, no.~10, pp. 3618--3631,
  Oct. 2012.

\bibitem{Devarajan2012}
R.~Devarajan, S.~Jha, U.~Phuyal, and V.~Bhargava, ``Energy-aware resource
  allocation for cooperative cellular network using multi-objective
  optimization approach,'' \emph{IEEE Transactions on Wireless Communications},
  vol.~11, no.~5, pp. 1797--1807, May 2012.

\bibitem{Cheung2013}
K.~T.~K. Cheung, S.~Yang, and L.~Hanzo, ``Achieving maximum energy-efficiency
  in multi-relay {OFDMA} cellular networks: A fractional programming
  approach,'' \emph{IEEE Transactions on Communications}, vol.~61, no.~7, pp.
  2746--2757, Jul. 2013.

\bibitem{Cheung2013a}
------, ``Maximizing energy-efficiency in multi-relay {OFDMA} cellular
  networks,'' in \emph{Proceedings of the IEEE Global Telecommunications
  Conference}, Atlanta, GA, USA, Dec. 2013.

\bibitem{Cheung2014}
------, ``Spectral and energy spectral efficiency optimization of joint
  transmit and receive beamforming based multi-relay {MIMO}-{OFDMA} cellular
  networks,'' \emph{submitted to IEEE Transactions on Wireless Communications}.

\bibitem{Gesbert2010}
D.~Gesbert, S.~Hanly, H.~Huang, S.~Shamai~Shitz, O.~Simeone, and W.~Yu,
  ``Multi-cell {MIMO} cooperative networks: A new look at interference,''
  \emph{IEEE Journal on Selected Areas in Communications}, vol.~28, no.~9, pp.
  1380--1408, Dec. 2010.

\bibitem{Dinkelbach1967}
W.~Dinkelbach, ``On nonlinear fractional programming,'' \emph{Management
  Science}, vol.~13, pp. 492--498, Mar. 1967.

\bibitem{Avriel1988}
M.~Avriel, W.~E. Diewert, S.~Schaible, and I.~Zang, \emph{Generalized
  Concavity}.\hskip 1em plus 0.5em minus 0.4em\relax Plenum Press, New York,
  NY, USA, 1988.

\bibitem{Boyd2004}
S.~Boyd and L.~Vandenberghe, \emph{Convex Optimization}.\hskip 1em plus 0.5em
  minus 0.4em\relax Cambridge University Press, New York, NY, USA, 2004.

\bibitem{Isheden2012}
C.~Isheden, Z.~Chong, E.~Jorswieck, and G.~Fettweis, ``Framework for link-level
  energy efficiency optimization with informed transmitter,'' \emph{IEEE
  Transactions on Wireless Communications}, vol.~11, no.~8, pp. 2946--2957,
  Aug. 2012.

\bibitem{Sung2010}
H.~Sung, S.-H. Park, K.-J. Lee, and I.~Lee, ``Linear precoder designs for
  ${K}$-user interference channels,'' \emph{IEEE Transactions on Wireless
  Communications}, vol.~9, no.~1, pp. 291--301, Jan. 2010.

\bibitem{Alexandropoulos2013}
G.~Alexandropoulos and C.~Papadias, ``A reconfigurable distributed algorithm
  for ${K}$-user {MIMO} interference networks,'' in \emph{IEEE International
  Conference on Communications}, Budapest, Hungary, Jun. 2013, pp. 3063--3067.

\bibitem{Ronasi2014}
K.~Ronasi, B.~Niu, V.~W. Wong, S.~Gopalakrishnan, and R.~Schober,
  ``Throughput-efficient scheduling and interference alignment for {MIMO}
  wireless systems,'' \emph{IEEE Transactions on Wireless Communications},
  vol.~13, no.~4, pp. 1779--1789, Apr. 2014.

\bibitem{Chen2014}
X.~Chen and C.~Yuen, ``Performance analysis and optimization for interference
  alignment over {MIMO} interference channels with limited feedback,''
  \emph{IEEE Transactions on Signal Processing}, vol.~62, no.~7, pp.
  1785--1795, Apr. 2014.

\bibitem{Palomar2006}
D.~Palomar and M.~Chiang, ``A tutorial on decomposition methods for network
  utility maximization,'' \emph{IEEE Journal on Selected Areas in
  Communications}, vol.~24, no.~8, pp. 1439--1451, Aug. 2006.

\bibitem{Raleigh1998}
G.~Raleigh and J.~Cioffi, ``Spatio-temporal coding for wireless
  communication,'' \emph{IEEE Transactions on Communications}, vol.~46, no.~3,
  pp. 357--366, Mar. 1998.

\bibitem{UlHassan2009}
N.~Ul~Hassan and M.~Assaad, ``Low complexity margin adaptive resource
  allocation in downlink {MIMO}-{OFDMA} system,'' \emph{IEEE Transactions on
  Wireless Communications}, vol.~8, no.~7, pp. 3365--3371, Jul. 2009.

\bibitem{Blum2003}
R.~Blum, ``{MIMO} capacity with interference,'' \emph{IEEE Journal on Selected
  Areas in Communications}, vol.~21, no.~5, pp. 793--801, Jun. 2003.

\bibitem{Hanzo2009}
L.~Hanzo, O.~Alamri, M.~El-Hajjar, and N.~Wu, \emph{Near-Capacity
  Multi-Functional {MIMO} Systems: Sphere-Packing, Iterative Detection and
  Cooperation}.\hskip 1em plus 0.5em minus 0.4em\relax Wiley-IEEE Press, 2009.

\bibitem{Auer2010}
G.~Auer, O.~Blume, V.~Giannini, I.~Godor, M.~A. Imran, Y.~Jading,
  E.~Katranaras, M.~Olsson, D.~Sabella, P.~Skillermark, and W.~Wajda, ``D2.3:
  Energy efficiency analysis of the reference systems, areas of improvements
  and target breakdown,'' \emph{{INFSO-ICT-247733 EARTH (Energy Aware Radio and
  NeTwork TecHnologies), Technical Report}}, Nov. 2010.

\bibitem{Bertsekas1999}
D.~P. Bertsekas, \emph{Nonlinear Programming}.\hskip 1em plus 0.5em minus
  0.4em\relax Athena Scientific, Belmont, MA, USA, 1999.

\bibitem{Yu2006}
W.~Yu and R.~Lui, ``Dual methods for nonconvex spectrum optimization of
  multicarrier systems,'' \emph{IEEE Transactions on Communications}, vol.~54,
  no.~7, pp. 1310--1322, Jul. 2006.

\bibitem{Ng2012c}
D.~Ng, E.~Lo, and R.~Schober, ``Dynamic resource allocation in {MIMO}-{OFDMA}
  systems with full-duplex and hybrid relaying,'' \emph{IEEE Transactions on
  Communications}, vol.~60, no.~5, pp. 1291--1304, May 2012.

\bibitem{Ng2012}
------, ``Energy-efficient resource allocation for secure {OFDMA} systems,''
  \emph{IEEE Transactions on Vehicular Technology}, vol.~61, no.~6, pp.
  2572--2585, Jul. 2012.

\bibitem{Seong2006}
K.~Seong, M.~Mohseni, and J.~Cioffi, ``Optimal resource allocation for {OFDMA}
  downlink systems,'' in \emph{Proceedings of the IEEE International Symposium
  on Information Theory}, Seattle, Washington, USA, Jul. 2006, pp. 1394--1398.

\bibitem{Arnold2010}
O.~Arnold, F.~Richter, G.~Fettweis, and O.~Blume, ``Power consumption modeling
  of different base station types in heterogeneous cellular networks,'' in
  \emph{Proceedings of the Future Network and Mobile Summit}, Florence, Italy,
  Jun. 2010.

\bibitem{3GPP_PL}
3GPP, ``{TR 36.814 V9.0.0:} further advancements for {E-UTRA}, physical layer
  aspects (release 9),'' Mar. 2010.

\end{thebibliography}

\begin{IEEEbiography}[{\includegraphics[scale=0.315]{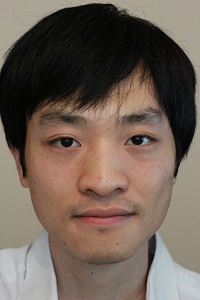}}]{Kent Tsz Kan Cheung}~(S'09) received his B.Eng. degree~(first-class honors) in electronic engineering from the Univeristy of Southampton, Southampton, U.K., in 2009. In 2015, he completed his Ph.D. degree in wireless communications at the same institution. He was a recipient of the EPSRC Industrial CASE award in 2009, and was involved with the Core 5 Green Radio project of the Virtual Centre of Excellence in Mobile and Personal Communications (Mobile VCE).

His research interests include energy-efficiency, multi-carrier MIMO communications, cooperative communications, resource allocation and optimization.\end{IEEEbiography}

\begin{IEEEbiography}[{\includegraphics[width=1in,height=1.25in,clip,keepaspectratio]{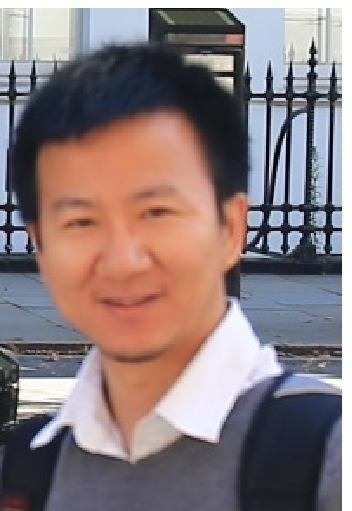}}] {Shaoshi Yang}
(S'09-M'13) received his B.Eng. degree in Information Engineering from Beijing University of Posts and Telecommunications (BUPT), Beijing, China in Jul. 2006, his first Ph.D. degree in Electronics and Electrical Engineering from University of Southampton, U.K. in Dec. 2013, and his second Ph.D. degree in Signal and Information Processing from BUPT in Mar. 2014. He is now working as a Postdoctoral Research Fellow in University of Southampton, U.K. From November 2008 to February 2009, he was an Intern Research Fellow with the Communications Technology Lab (CTL), Intel Labs, Beijing, China, where he focused on Channel Quality Indicator Channel (CQICH) design for mobile WiMAX (802.16m) standard. His research interests include MIMO signal processing, green radio, heterogeneous networks, cross-layer interference management, convex optimization and its applications. He has published in excess of 30 research papers on IEEE journals and conferences.

Shaoshi has received a number of academic and research awards, including the prestigious Dean's Award for Early Career Research Excellence at University of Southampton, the PMC-Sierra Telecommunications Technology Paper Award at BUPT, the Electronics and Computer Science (ECS) Scholarship of University of Southampton, and the Best PhD Thesis Award of BUPT. He is a member of IEEE/IET, and a junior member of Isaac Newton Institute for Mathematical Sciences, Cambridge University, U.K. He also serves as a TPC member of several major IEEE conferences, including \textit{IEEE ICC, GLOBECOM, PIMRC, ICCVE, HPCC}, and as a Guest Associate Editor of \textit{IEEE Journal on Selected Areas in Communications.} (https://sites.google.com/site/shaoshiyang/)
\end{IEEEbiography}

\begin{IEEEbiography}[{\includegraphics[width=1in,height=1.25in,clip,keepaspectratio]{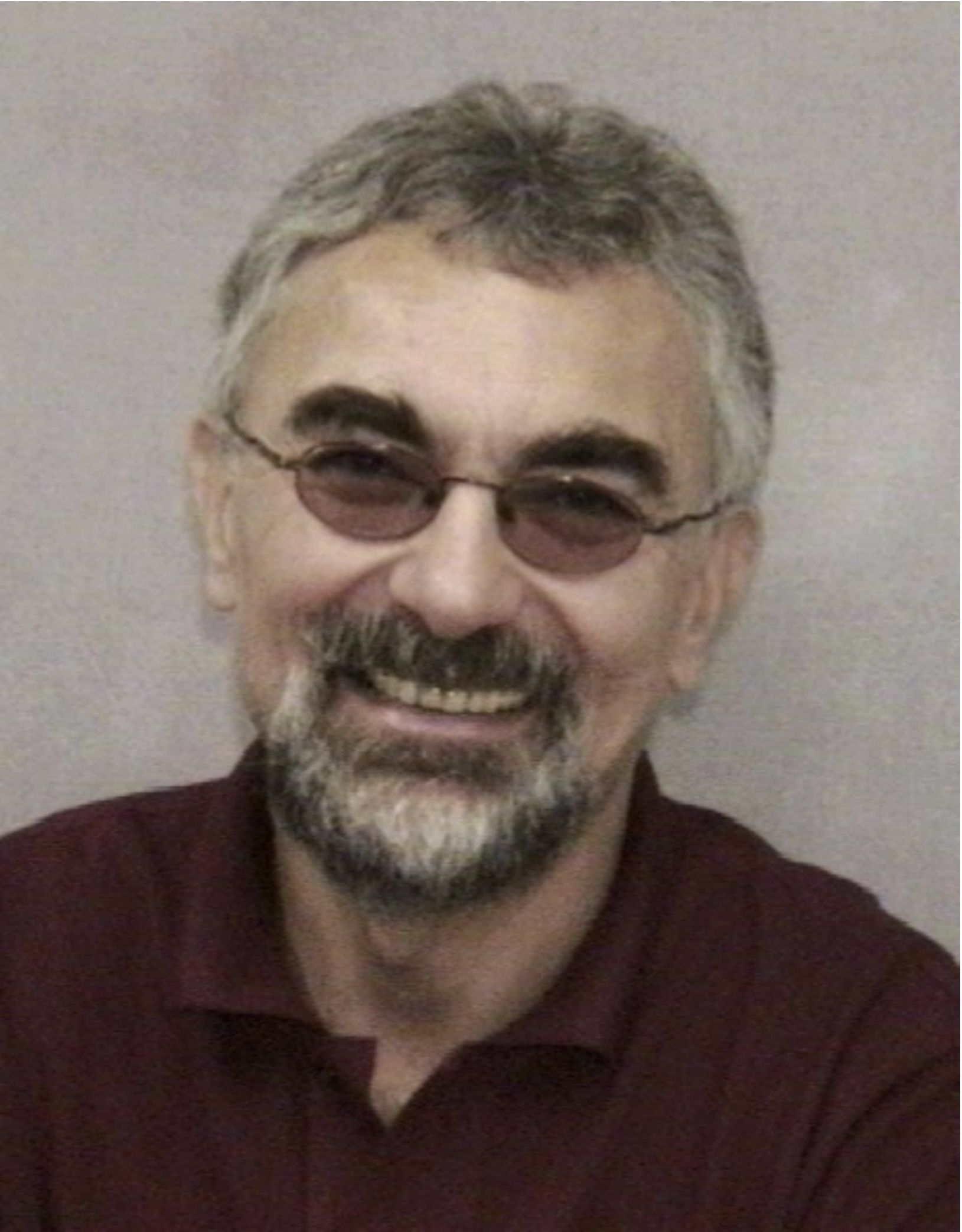}}] {Lajos Hanzo}~(http://www.cspc.ecs.soton.ac.uk)
FREng, FIEEE,
FIET, Fellow of EURASIP, DSc received his degree in electronics in
1976 and his doctorate in 1983. In 2009 he was awarded the honorary
doctorate ``Doctor Honoris Causa'' by the Technical University of
Budapest.

During his 35-year career in telecommunications he has held
various research and academic posts in Hungary, Germany and the
UK. Since 1986 he has been with the School of Electronics and Computer
Science, University of Southampton, UK, where he holds the chair in
telecommunications.  He has successfully supervised 80 PhD students,
co-authored 20 John Wiley/IEEE Press books on mobile radio
communications totalling in excess of 10 000 pages, published 1500+
research entries at IEEE Xplore, acted both as TPC and General Chair
of IEEE conferences, presented keynote lectures and has been awarded a
number of distinctions. Currently he is directing a 60-strong
academic research team, working on a range of research projects in the
field of wireless multimedia communications sponsored by industry, the
Engineering and Physical Sciences Research Council (EPSRC) UK, the
European IST Programme and the Mobile Virtual Centre of Excellence
(VCE), UK.

He is an enthusiastic supporter of industrial and academic
liaison and he offers a range of industrial courses.  He is also a
Governor of the IEEE VTS.  During 2008 - 2012 he was the
Editor-in-Chief of the IEEE Press and a Chaired Professor also at
Tsinghua University, Beijing.  His research is funded by the
European Research Council's Senior Research Fellow Grant.  For further
information on research in progress and associated publications please
refer to http://www-mobile.ecs.soton.ac.uk.
\end{IEEEbiography} 

\end{document}